\documentclass[%
 reprint,
 amsmath,amssymb,
 aps,
prb,
 superscriptaddress
]{revtex4-2}

\usepackage{graphicx}
\usepackage{dcolumn}
\usepackage{bm}
\usepackage{color}

\begin{document}

\preprint{APS/123-QED}

\title{Non-Bloch bands in two-dimensional non-Hermitian systems}

\author{Kazuki Yokomizo}
\affiliation{Department of Physics, The University of Tokyo, 7-3-1 Hongo, Bunkyo-ku, Tokyo, 113-0033, Japan}
\author{Shuichi Murakami}%
\affiliation{Department of Physics, Tokyo Institute of Technology, 2-12-1 Ookayama, Meguro-ku, Tokyo, 152-8551, Japan}%




%
\begin{abstract}
The non-Bloch band theory can describe energy bands in a one-dimensional (1D) non-Hermitian system. On the other hand, whether the non-Bloch band theory can be extended to higher-dimensional non-Hermitian systems is nontrivial. In this work, we construct the non-Bloch band theory in two classes of two-dimensional non-Hermitian systems, by reducing the problem to that for a 1D non-Hermitian model. In these classes of systems, we get the generalized Brillouin zone for a complex wavevector and investigate topological properties. In the model of the non-Hermitian Chern insulator, as an example, we show the bulk-edge correspondence between the Chern number defined from the generalized Brillouin zone and the appearance of the edge states.
\end{abstract}
\pacs{Valid PACS appear here}
\maketitle
%
%

\section{\label{sec1}Introduction}
A non-Hermitian Hamiltonian effectively describes a nonequilibrium system~\cite{Ashida2020}. In recent studies, the non-Hermitian skin effect, i.e., the localization of bulk eigenstates, plays a crucial role because it causes remarkable phenomena~\cite{Yao2018,Longhi2019,Song2019,Okuma2020,Zhang2020,Mcdonald2020,Yi2020,Li2020,Yokomizo2021v2,Longhi2022,Longhi2022v2,Brandenbourger2019,Xiao2020,Weidemann2020,Helbig2020,Hofmann2020,Ghatak2020,XZhang2021,Chen2021,Zhang2021,Wang2022,Liang2022,Liu2022}. In particular, with the non-Hermitian skin effect, energy eigenvalues under an open boundary condition and those under a periodic boundary condition are different. Such phenomena associated with the non-Hermitian skin effect are an obstacle to investigate properties of a non-Hermitian system. For example, the bulk-edge correspondence between a topological invariant defined from the conventional Bloch wavevector and existence of topological edge states seems to be violated within the conventional definition of a topological invariant.

Recently, the non-Bloch band theory was proposed~\cite{Yao2018,Yokomizo2019,Yokomizo2020,Yokomizo2021,Yokomizo2022}. It was shown that in a one-dimensional (1D) spatially periodic non-Hermitian system without disorder, the generalized Brillouin zone $\beta=e^{ik}$ for the complex Bloch wavenumber $k$ becomes different from that in a Hermitian system, and it reproduces energy eigenvalues under an open boundary condition. The non-Bloch band theory gives the condition for the generalized Brillouin zone and is applicable to the system even with long-range hopping amplitudes and on-site potentials. Importantly, a topological invariant defined from the generalized Brillouin zone restores the bulk-edge correspondence. Additionally, some features of the generalized Brillouin zone lead to unique phenomena, such as appearance of a topological semimetal phase with exceptional points~\cite{Yokomizo2020v2}. Thus, the non-Bloch band theory is a powerful tool for studies on many properties in a non-Hermitian system. However, there is a lack of studies on a two-dimensional (2D) non-Hermitian system in terms of the non-Bloch band theory.

There have been some studies on topological phases and boundary states in a 2D non-Hermitian system in terms of the conventional Bloch band theory. For example, some previous works investigated how non-Hermiticity affects the bulk-edge correspondence~\cite{Leykam2017,Shen2018,Kawabata2018,Ezawa2019,Luo2019,Wu2019,Wu2020,Xue2020,Ao2020,Song2020,Xie2021,Teo2022}. Furthermore, Refs.~\cite{Philip2018,Chen2018,Hirsbrunner2019,Groenendijk2021} calculated the Hall conductance associated with the topological edge states in the model of the non-Hermitian Chern insulator. In recent years, non-Hermitian topology in a higher-dimensional system has been attracting much attention. In theory, Refs.~\cite{Lee2019,Kawabata2020,Okugawa2020,Fu2021,Zhang2022,Li2022,Zhu2022,Fang2022} proposed the higher-order topological non-Hermitian skin effect, in which eigenstates are localized at corners of the system, and in experiment, it has been observed~\cite{Palacios2021,Zou2021,Xu2022,Shang2022,Wu2022}. We emphasize that the topological invariant predicting the higher-order topological non-Hermitian skin effect is defined from the real Bloch wavevector. However, as we learned from a 1D non-Hermitian system, we cannot investigate physics of a 2D non-Hermitian system with an open boundary condition in terms of the real Bloch wavevector. Therefore, it is necessary to construct the non-Bloch band theory in a 2D non-Hermitian system.

So far, the non-Bloch bands in some 2D non-Hermitian models were investigated~\cite{Yao2018v2,Liu2019,Yu2021,Lin2021,Xiao2022,Bartlett2023,Jiang2022}. Nevertheless, it is unclear whether the non-Bloch band theory in general 2D non-Hermitian systems can be constructed. In this work, we construct the non-Bloch band theory in two classes of 2D non-Hermitian systems, where we can reduce the problem to that of a 1D non-Hermitian model. One class of systems has specific symmetries, such as a mirror symmetry suppressing the non-Hermitian skin effect in one direction. The other class is decoupled into two 1D non-Hermitian systems due to a special form of the Hamiltonian. These properties indicate that the eigenstates in the bulk are written as linear combination of a few plane waves. Thereby, the condition for the generalized Brillouin zone can be obtained. In this paper, we exemplify three systems, e.g., the model of the non-Hermitian Chern insulator, the non-Hermitian Benalcazar–Bernevig–Hughes (BBH) model, and the Okugawa-Takahashi-Yokomizo (OTY) model. We show that the generalized Brillouin zone reproduces the energy eigenvalues in these models with a full open boundary condition. Furthermore, we investigate topological properties in terms of the generalized Brillouin zone.

This paper is organized as follows: In Sec.~\ref{sec2}, we introduce our 2D non-Hermitian tight-binding model and propose two classes of systems where we can apply the non-Bloch band theory. Then, we can get the condition for the generalized Brillouin zone. In Sec.~\ref{sec3}, we analyze the three 2D non-Hermitian models in terms of the non-Bloch band theory. In Sec.~\ref{sec4}, we discuss the 2D non-Bloch band theory proposed in this work from the viewpoint of the formation of standing wave and comment on difficulty to construct the non-Bloch band theory in general 2D non-Hermitian systems. Finally, in Sec.~\ref{sec5}, we summarize our result.

%
%

\section{\label{sec2}Model}
\begin{figure}[]
\includegraphics[width=6cm]{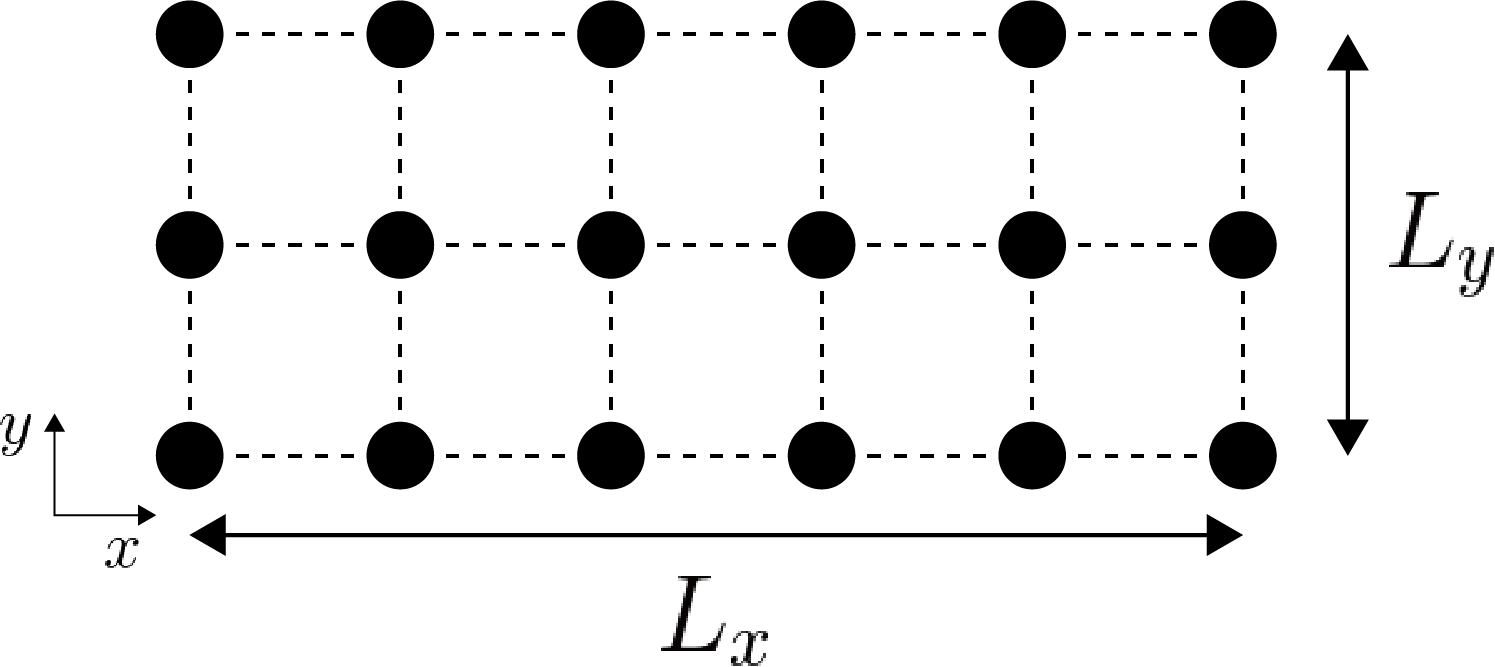}
\caption{\label{fig1}Two-dimensional tight-binding system on a square lattice. The geometry of the system is a rectangle, where the lengths in the $x$ direction and in the $y$ direction are $L_x$ and $L_y$, respectively. The black circles express the unit cells at the lattice point ${\bm r}=\left(n_x,n_y\right)~\left(n_x=1,\dots,L_x,n_y=1,\dots,L_y\right)$.}
\end{figure}
First of all, we introduce a 2D non-Hermitian tight-binding system. Throughout this paper, the system lies on a square lattice, and the geometry of the system is a rectangle, as shown in Fig.~\ref{fig1}. The real-space Hamiltonian of our system is given by
\begin{eqnarray}
H&=&\sum_{{\bm r}}\sum_{\mu,\nu=1}^q\left(\sum_{i=-N_x}^{N_x}t_{i,\mu\nu}^{\left(x\right)}c_{{\bm r}+i\hat{\bm x},\mu}^\dag c_{{\bm r},\nu}\right. \nonumber\\
&&\left.+\sum_{i=-N_y}^{N_y}t_{i,\mu\nu}^{\left(y\right)}c_{{\bm r}+i\hat{\bm y},\mu}^\dag c_{{\bm r},\nu}\right),
\label{eq1}
\end{eqnarray}
where $c_{{\bm r},\mu}^\dag$ is a creation operator, ${\bm r}=\left(n_x,n_y\right)~\left(n_x=1,\dots,L_x,n_y=1,\dots,L_y\right)$ is a lattice point, $q$ is the number of internal degrees of freedom in a unit cell, and $\hat{\bm x}$ and $\hat{\bm y}$ are unit vectors in the $x$ direction and in the $y$ direction, respectively. The particles hop up to the $N_x$th and $N_y$th nearest-neighbor unit cells along the $x$ and $y$ directions, respectively. We note that when either $t_{i,\nu\mu}^{\left(x\right)}\neq\left(t_{-i,\mu\nu}^{\left(x\right)}\right)^\ast$ or $t_{i,\nu\mu}^{\left(y\right)}\neq\left(t_{-i,\mu\nu}^{\left(y\right)}\right)^\ast$ is satisfied, the system becomes non-Hermitian. Now, the real-space eigen-equation is written as
\begin{equation}
H\left|{\psi}\right\rangle=E\left|\psi\right\rangle,
\label{eq2}
\end{equation}
where the eigenstates are given by
\begin{equation}
\left|\psi\right\rangle=\left(\dots,\psi_{\left(n_x,n_y\right),1},\dots,\psi_{\left(n_x,n_y,1\right),q},\dots\right)^{\rm T}.
\label{eq3}
\end{equation}

Next, we describe a way to construct the non-Bloch band theory in our system. In the case of $q=1$, it is straightforward to construct the non-Bloch band theory in the system. This is because the system can be decomposed into two 1D non-Hermitian models. In Appendix~\ref{secA}, we derive the condition for the generalized Brillouin zone in this case. On the other hand, in the case of $q\geq2$, we find that when the system satisfies some conditions, one can calculate the generalized Brillouin zone in terms of the conventional non-Bloch band theory. In the following, we explain the two cases where the description of the non-Bloch bands is possible.

%
%

\subsection{\label{sec2-1}Case A}
In a 1D non-Hermitian system, when the energy of the eigenstate with the Bloch wavenumber $+k$ and that of the eigenstate with $-k$ are degenerate, the non-Hermitian skin effect is suppressed because $k$ becomes real~\cite{Kawabata2019,Yi2020,Okugawa2021}. The degeneracy leading to the suppression of the non-Hermitian skin effect comes from symmetries, e.g., a global symmetry, such as a PT symmetry, and a crystalline symmetry, such as a mirror symmetry. Importantly, we here propose that such suppression is realized also in a 2D non-Hermitian system. For example, when the Bloch Hamiltonian $H\left(k_x,k_y\right)$ satisfies
\begin{equation}
U^{-1}H^{\rm T}\left(k_x,-k_y\right)U=H\left(k_x,k_y\right),
\label{eq4}
\end{equation}
where $U$ is a unitary matrix satisfying $U^2=1$, we can show that the non-Hermitian skin effect in the $y$ direction is suppressed. In fact, when Eq.~(\ref{eq4}) is satisfied, the wave with the Bloch wavevector $\left(k_x,k_y\right)$ is reflected to that with $\left(k_x,-k_y\right)$ at the boundary of the system parallel to the $x$ direction. Then, the condition for the formation of standing wave means that the imaginary parts of the wavevector for the incident and reflected waves are equal. Therefore, we can obtain ${\rm Im}\left(k_y\right)={\rm Im}\left(-k_y\right)$, meaning that $k_y$ is real, and the non-Hermitian skin effect in the $y$ direction is suppressed.

In this case, the bulk eigenstates extend over the $y$ direction because the Bloch wavenumber $k_y$ becomes real. Hence, in the limit of a large system size, the asymptotic behavior of the bulk eigenstates in the cylinder geometry, where a periodic boundary condition in the $y$ direction is imposed to the system, matches that in the full-open geometry. Therefore, the energy bands in the cylinder geometry asymptotically reproduces the energy levels in a finite open plane. Importantly, since the cylinder geometry can be regarded as a pseudo-1D system, one can calculate the energy bands of this system in terms of the conventional non-Bloch band theory. Thus, we can construct the non-Bloch band theory in the system.

Based on the above concept, we establish the non-Bloch band theory in the 2D non-Hermitian systems without the non-Hermitian skin effect in the $y$ direction. With Eq.~(\ref{eq4}), by assuming that the Bloch wavenumber $k_y$ is real, we take
\begin{equation}
\psi_{{\bm r},\mu}=\sum_j\left(\beta_{x,j}\right)^{n_x}e^{ik_yn_y}\phi_\mu^{\left(j\right)}
\label{eq5}
\end{equation}
as an ansatz for the eigen-equation in the cylinder geometry. Here, we define the non-Bloch matrix as
\begin{equation}
\left[{\cal H}\left(\beta_x,k_y\right)\right]_{\mu\nu}=\sum_{m=-N_x}^{N_x}t_{m,\mu\nu}^{\left(x\right)}\left(\beta_x\right)^m+\sum_{m=-N_y}^{N_y}t_{m,\mu\nu}^{\left(y\right)}e^{ik_ym}.
\label{eq6}
\end{equation}
Then, in Eq.~(\ref{eq5}), $\beta_x=\beta_{x,j}$ is the solution of the characteristic equation
\begin{equation}
\det\left[{\cal H}\left(\beta_x,k_y\right)-E\right]=0.
\label{eq7}
\end{equation}
We note that Eq.~(\ref{eq7}) is an algebraic equation for $\beta_x$ of $2qN_x$th degree. Therefore, the condition for the generalized Brillouin zone is given by
\begin{equation}
\left|\beta_{x,qN_x}\right|=\left|\beta_{x,qN_x+1}\right|,
\label{eq8}
\end{equation}
where the solutions of Eq.~(\ref{eq7}) are numbered as
\begin{equation}
\left|\beta_{x,1}\right|\leq\dots\leq\left|\beta_{x,2qN_x}\right|.
\label{eq9}
\end{equation}
For a given real value of $k_y$, we can obtain trajectories of $\beta_{x,qN_x}$ and $\beta_{x,qN_x+1}$ with Eq.~(\ref{eq8}) forming loops on the complex plane. Thus, by changing the real value of $k_y$, the surface formed by $\left(\beta_x,\beta_y\right)~\left(\beta_y\equiv e^{ik_y},k_y\in{\mathbb R}\right)$ is the generalized Brillouin zone. Finally, from the generalized Brillouin zone, we can calculate the energy bands by using Eq.~(\ref{eq7}). In this work, we show that it matches the energy levels in a finite open plane. This means that the non-Hermitian skin effect is indeed suppressed in the $y$ direction in this case.

%
%

\subsection{\label{sec2-2}Case B}
Next, we focus on the case that the characteristic equation of the Bloch Hamiltonian $H\left(k_x,k_y\right)$ is written in the form of separation of variables. In this case, the motion of the bulk eigenstates in the $x$ direction is completely decoupled from that in the $y$ direction. Namely, the 2D non-Hermitian systems can be regarded as two 1D non-Hermitian systems. In the following, we assume that the characteristic equation is given by
\begin{equation}
\det\left[H\left(k_x,k_y\right)-E\right]=R_x\left(e^{ik_x},E\right)+R_y\left(e^{ik_y}\right)=0.
\label{eq10}
\end{equation}
Then, by applying the conventional non-Bloch band theory to the 1D non-Hermitian systems, we expect that one can get the energy bands in the 2D non-Hermitian system, and indeed, we will show it in this work.

Now, we derive the condition for the generalized Brillouin zone to calculate the energy bands. With Eq.~(\ref{eq10}), we have
\begin{equation}
\psi_{{\bm r},\mu}=\sum_{j_x,j_y}\left(\beta_{x,j_x}\right)^{n_x}\left(\beta_{y,j_y}\right)^{n_y}\phi_\mu^{\left(j_x,j_y\right)}
\label{eq11}
\end{equation}
as an ansatz for Eq.~(\ref{eq2}). Here, $\beta_x=\beta_{x.j_x}$ and $\beta_y=\beta_{y.j_y}$ in Eq.~(\ref{eq11}) are the solutions of the equation $R_x\left(\beta_x,E\right)=\lambda$ and the equation $R_y\left(\beta_y\right)=-\lambda$, respectively, where $\lambda$ is an arbitrary constant. We note that these equations are algebraic equations of $2qN_x$th degree and $2qN_y$th degree, respectively. Thus, we can get the generalized Brillouin zone spanned by $\left(\beta_x,\beta_y\right)$ satisfying
\begin{equation}
\left|\beta_{a,qN_a}\right|=\left|\beta_{a,qN_a+1}\right|,
\label{eq12}
\end{equation}
where the solutions of $R_x\left(\beta_x,E\right)=\lambda$ and $R_y\left(\beta_y\right)=-\lambda$ are numbered as
\begin{equation}
\left|\beta_{a,1}\right|\leq\dots\leq\left|\beta_{a,2qN_a}\right|
\label{eq13}
\end{equation}
for $a=x,y$. Finally, the energy bands are obtained from the characteristic equation of the non-Bloch matrix
\begin{equation}
\left[{\cal H}\left(\beta_x,\beta_y\right)\right]_{\mu\nu}=\sum_{m=-N_x}^{N_x}t_{m,\mu\nu}^{\left(x\right)}\left(\beta_x\right)^m+\sum_{m=-N_y}^{N_y}t_{m,\mu\nu}^{\left(y\right)}\left(\beta_y\right)^m.
\label{eq14}
\end{equation}
We note that unlike Case A, the non-Hermitian skin effect appear in both directions.

%
%

\section{\label{sec3}Examples}
In this section, we calculate the generalized Brillouin zone and the energy bands in some examples, e.g., the model of the non-Hermitian Chern insulator, the non-Hermitian BBH model, and the OTY model. The first model is in Case A, and the second and third models are in Case B, as discussed in Sec.~\ref{sec2}. Throughout this calculation, we show that the non-Bloch band theory in the Case A and Case B indeed reproduces the energy levels in a finite open plane. Furthermore, we study topological properties in theses systems in terms of the generalized Brillouin zone.

%
%

\subsection{\label{sec3-1}Non-Hermitian Chern insulator}
\begin{figure}[]
\includegraphics[width=8.5cm]{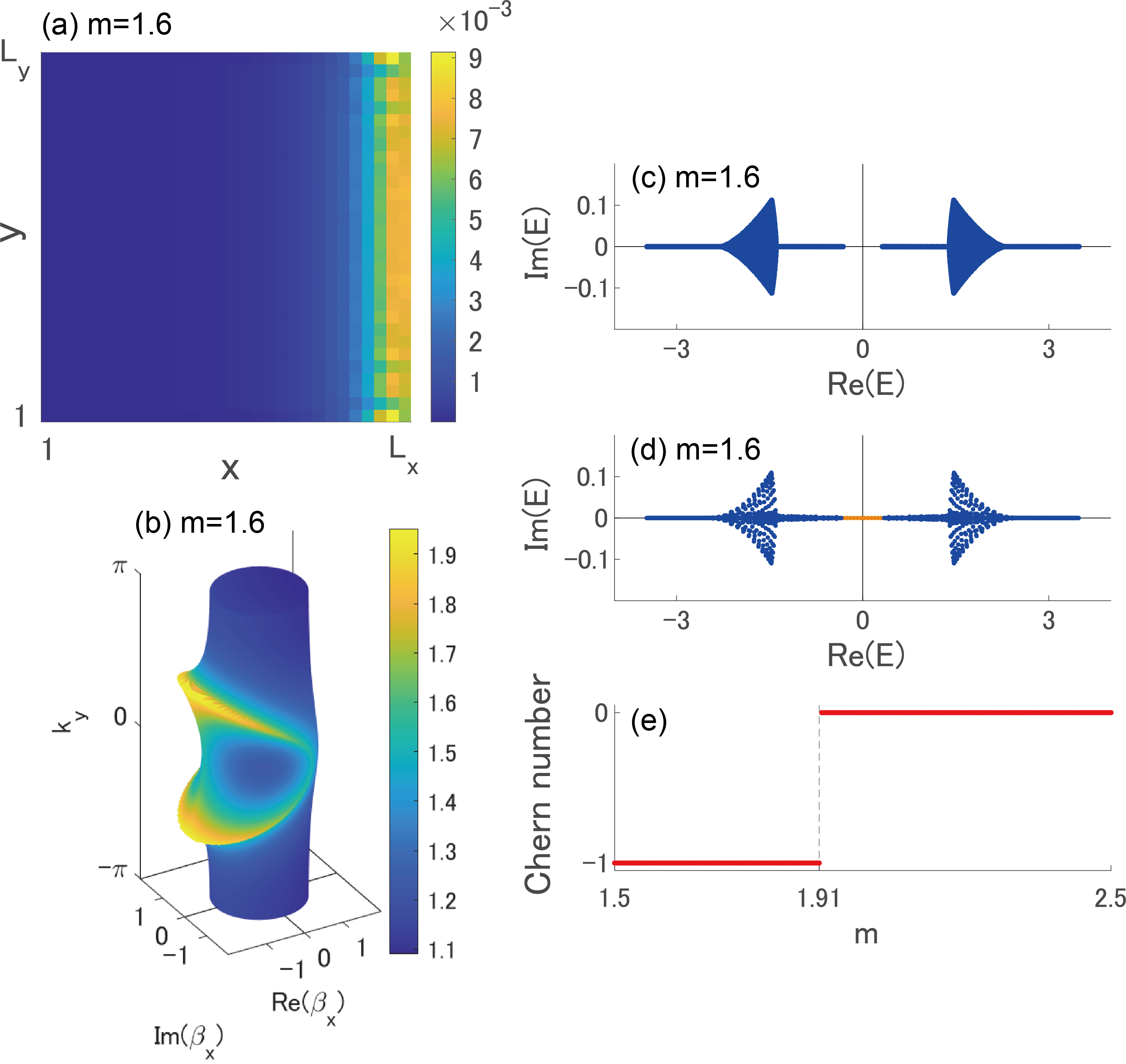}
\caption{\label{fig2}Eigenstates, generalized Brillouin zone, energy eigenvalue, and Chern number in the model of the non-Hermitian Chern insulator. (a) Bulk eigenstates localized at one edge. (b) Generalized Brillouin zone spanned by $\left(\beta_x,\beta_y\right)~\left(\beta_y\equiv e^{ik_y},k_y\in{\mathbb R}\right)$. The color bar means the absolute value of $\beta_x$. (c) Energy bands calculated from the generalized Brillouin zone. (d) Energy level in the system with $L_x=L_y=30$. We show the topological edge states in orange. (e) Chern number for the band with ${\rm Re}\left(E\right)<0$ computed by using the Fukui-Hatsugai-Suzuki method. The Chern number changes its value at $m=1.91$. In panels (a)-(e), the values of the system parameters are set to be $t_x=0.9,t_y=1,v_x=1.2,v_y=0.8,\gamma_x=0.2$, and $\gamma_y=0$, and in panels (a)-(d), $m=1.6$.}
\end{figure}
First of all, we focus on the model of a non-Hermitian Chern insulator investigated in Ref.~\cite{Yao2018v2}. The real-space Hamiltonian of this system is given by
\begin{eqnarray}
H&=&\sum_{\bm r}\left({\bm c}_{{\bm r}+\hat{\bm x}}^\dag T_x^\dag{\bm c}_{{\bm r}}+{\bm c}_{{\bm r}}^\dag T_x{\bm c}_{{\bm r}+\hat{\bm x}}\right. \nonumber\\
&&\left.+{\bm c}_{{\bm r}+\hat{\bm y}}^\dag T_y^\dag{\bm c}_{{\bm r}}+{\bm c}_{{\bm r}}^\dag T_y{\bm c}_{{\bm r}+\hat{\bm y}}+{\bm r}_{{\bm n}}^\dag M{\bm c}_{{\bm r}}\right),
\label{eq15}
\end{eqnarray}
where
\begin{eqnarray}
T_x&=&\frac{1}{2}\left( \begin{array}{cc}
-t_x  & -iv_x \vspace{3pt}\\
-iv_x & t_x
\end{array}\right),~T_y=\frac{1}{2}\left( \begin{array}{cc}
-t_y & -v_y \vspace{3pt}\\
v_y  & t_y
\end{array}\right) \nonumber\\
M&=&\left( \begin{array}{cc}
m                  & i\gamma_x+\gamma_y \vspace{3pt}\\
i\gamma_x-\gamma_y & -m
\end{array}\right),
\label{eq16}
\end{eqnarray}
and ${\bm c}_{\bm r}^\dag=\left(c_{{\bm r},{\rm A}}^\dag,c_{{\bm r},{\rm B}}^\dag\right)$ are the creation operators of particles on sublattices A and B, respectively. For simplicity, all the parameters are set to be real. We note that this system is in the case of $N_x=1,N_y=1$, and $q=2$ in Eq.~(\ref{eq1}). The real-space eigen-equation is written as
\begin{equation}
T_x{\bm\psi}_{{\bm r}+\hat{\bm x}}+T_x^\dag{\bm\psi}_{{\bm r}-\hat{\bm x}}+T_y{\bm\psi}_{{\bm r}+\hat{\bm y}}+T_y^\dag{\bm\psi}_{{\bm r}-\hat{\bm y}}+M{\bm\psi}_{\bm r}=E{\bm\psi}_{\bm r},
\label{eq17}
\end{equation}
where ${\bm\psi}_{\bm r}=\left(\psi_{{\bm r},{\rm A}},\psi_{{\bm r},{\rm B}}\right)^{\rm T}$. In the following, we investigate the case of $\gamma_y=0$. Then, the Bloch Hamiltonian $H\left(k_x,k_y\right)$ of the system has the symmetry written as Eq.~(\ref{eq4}) with $U=1$. Hence, this system is in Case A as discussed in Sec.~\ref{sec2-1}, and we expect that the non-Hermitian skin effect is suppressed along the $y$ direction. Indeed, one can see from Fig.~\ref{fig2}(a) that all the bulk eigenstates are localized at one edge. By following the discussion in Sec.~\ref{sec2-1}, we can get the generalized Brillouin zone and the energy bands.

In the model of the non-Hermitian Chern insulator, the non-Bloch matrix is given by
\begin{equation}
{\cal H}\left(\beta_x,k_y\right)=\beta_xT_x+\beta_x^{-1}T_x^\dag+e^{ik_y}T_y+e^{-ik_y}T_y^\dag+M.
\label{eq18}
\end{equation}
Since the characteristic equation $\det\left[{\cal H}\left(\beta_x,k_y\right)-E\right]=0$ is a quartic equation for $\beta_x$, the condition for the generalized Brillouin zone is given by
\begin{equation}
\left|\beta_{x,2}\right|=\left|\beta_{x,3}\right|,
\label{eq19}
\end{equation}
where the solutions of the characteristic equation satisfy Eq.~(\ref{eq9}). As an example, we show the generalized Brillouin zone formed from the complex Bloch wavenumber $k_x$ and the real Bloch wavenumber $k_y$ in Fig.~\ref{fig2}(b). Furthermore, the energy bands are calculated from the generalized Brillouin zone as shown in Fig.~\ref{fig2}(c). This model has two bands, one with ${\rm Re}\left(E\right)>0$ and the other with ${\rm Re}\left(E\right)<0$. We also calculate the energy levels of the system on a finite open plane as shown in Fig.~\ref{fig2}(d). Comparing Fig,~\ref{fig2}(c) with Fig.~\ref{fig2}(d), we confirm that our non-Bloch bands reproduce the energy levels in a finite open plane.

Next, we investigate a topological aspect of the model of the non-Hermitian Chern insulator in terms of the Chern number defined from the generalized Brillouin zone. Let $\left|u_{R,n}\left({\bm k}\right)\right\rangle$ and $\left\langle u_{L,n}\left({\bm k}\right)\right|$ denote the right and left eigenvectors of Eq.~(\ref{eq18}), respectively, where ${\bm k}$ is the complex Bloch wavevector with $k_x$ defined by $e^{ik_x}\equiv\beta_x$. Here, $n$ expresses a band index, and we set $n=+$ and $n=-$ to represent the energy bands with ${\rm Re}\left(E\right)>0$ and ${\rm Re}\left(E\right)<0$, respectively. The right and left eigenvectors satisfy $\left\langle u_{L,m}\left({\bm k}\right)\middle|u_{R,n}\left({\bm k}\right)\right\rangle=\delta_{m,n}$. Then, we define the Chern number as
\begin{eqnarray}
C_n&=&\frac{1}{2\pi i}\int_{T_\beta}d{\bm k}{\cal B}_n\left({\bm k}\right), \label{eq20}\\
{\cal B}_n\left({\bm k}\right)&=&\left\langle\frac{\partial u_{L,n}\left({\bm k}\right)}{\partial k_x}\middle|\frac{\partial u_{R,n}\left({\bm k}\right)}{\partial k_y}\right\rangle \nonumber\\
&&-\left\langle\frac{\partial u_{L,n}\left({\bm k}\right)}{\partial k_y}\middle|\frac{\partial u_{R,n}\left({\bm k}\right)}{\partial k_x}\right\rangle, \label{eq21}
\end{eqnarray}
where ${\cal B}_n\left({\bm k}\right)$ is the Berry curvature. The definition of Eq.~(\ref{eq20}) means that the Chern number is defined as an integral of the Berry curvature over the generalized Brillouin zone denoted by $T_\beta$. Now, we compute the Chern number with $n=-$ by using the Fukui-Hatsugai-Suzuki method~\cite{Fukui2005}. As shown in Fig.~\ref{fig2}(e), the Chern number takes $-1$ in some region of the parameter $m$. Importantly, the nonzero Chern number corresponds to appearance of the topological edge states as we discuss next. Indeed, the topological edge states shown in orange appear in Fig.~\ref{fig2}(d). Thus, the non-Bloch band theory shows the bulk-edge correspondence in the model of the non-Hermitian Chern insulator. We note that the gap-closing point at $m=1.91$ is not an exceptional point because the non-Bloch matrix has full rank. It was shown by Ref.~ \cite{Bartlett2023} that the gapless phases with exceptional points can appear as an intermediate phase between the trivial phase and the topological phase in some parameter regions.

In general, the Chern number (\ref{eq20}) predicts existence of topological edge states. In the following, we focus on a non-Hermitian system with a line gap in which energy bands are separated into two sets, set 1 and set 2, by a line on the complex energy plane. First, we note that the sum of Eq.~(\ref{eq20}) over all the bands belonging to the set 1 is quantized, which is shown in Appendix~\ref{secB}. According to Ref.~\cite{Kawabata2019}, a non-Hermitian system with a line gap can be transformed to a Hermitian system without closing the gap. The transformation corresponds to the deformation from the generalized Brillouin zone to the conventional Brillouin zone. Then, the values of Eq.~(\ref{eq20}) remain quantized under the deformation. Hence, after transforming the system, Eq.~(\ref{eq20}) becomes the Chern number of the Hermitian system, defined by the conventional Brillouin zone. In the Hermitian system, the bulk-edge correspondence for the Chern number is established. Importantly, since the bulk gap remains open before and after the transformation, the topological properties in the non-Hermitian system and the Hermitian system are common. Therefore, the nonzero value of Eq.~(\ref{eq20}) corresponds to appearance of topological edge states across the gap between set 1 and set 2. Thus, we have shown the bulk-edge correspondence for Eq.~(\ref{eq20}).

Finally, we comment on the relationship between the present work and Ref.~\cite{Yao2018v2}. This previous work investigated the model of the non-Hermitian Chen insulator in terms of a low-energy continuum model in which the Bloch Hamiltonian of Eq.~(\ref{eq15}) is expanded up to second order of the Bloch wavevector around $\left(k_x,k_y\right)=\left(0,0\right)$. In the low-energy continuum model, one can obtain an approximated value of the complex-valued Bloch wavevector only near $\left(k_x,k_y\right)=\left(0,0\right)$. On the other hand, in this work, we show a way to get the full generalized Brillouin zone in the original lattice model. We note that near $\left(k_x,k_y\right)=\left(0,0\right)$, our generalized Brillouin zone reproduces the approximated result in the previous work. In particular, the previous work proposed that the band gap closes at
\begin{equation}
m_c\simeq t_x+t_y+\frac{t_x\gamma_x^2}{2v_x^2}+\frac{t_y\gamma_y^2}{2v_y^2},
\label{eq22}
\end{equation}
and the Chern number changes its value there. With the values of the system parameters in our calculation, Eq.~(\ref{eq22}) gives $m_c\simeq1.91$, and it matches the value of $m$ at which the Chern number changes in Fig.~\ref{fig2}(e).

%
%

\subsection{\label{sec3-2}Non-Hermitian Benalcazar–Bernevig–Hughes model}
Next, we study the non-Hermitian BBH model with the asymmetric intercell hopping amplitude and the next-nearest neighbor hopping amplitude by following Ref.~\cite{Okugawa2019}, which has studied the second-order topological phase in a Hermitian system with a chiral symmetry. The real-space Hamiltonian of this system is given by
\begin{eqnarray}
H&=&\sum_{\bm r}\left({\bm c}_{{\bm r}+\hat{\bm x}}^\dag T_x^\dag{\bm c}_{{\bm r}}+{\bm c}_{{\bm r}}^\dag T_x{\bm c}_{{\bm r}+\hat{\bm x}}\right. \nonumber\\
&&\left.+{\bm c}_{{\bm r}+\hat{\bm y}}^\dag T_y^\dag{\bm c}_{{\bm r}}+{\bm c}_{{\bm r}}^\dag T_y{\bm c}_{{\bm r}+\hat{\bm y}}+{\bm c}_{{\bm r}}^\dag M{\bm c}_{{\bm r}}\right),
\label{eq23}
\end{eqnarray}
where
\begin{eqnarray}
T_x&=&\left( \begin{array}{cccc}
0     & 0      & t_3^x & 0      \vspace{3pt}\\
0     & 0      & 0     & -t_3^x \vspace{3pt}\\
t_2^x & 0      & 0     & 0      \vspace{3pt}\\
0     & -t_2^x & 0     & 0
\end{array}\right),~T_y=\left( \begin{array}{cccc}
0     & t_3^y & 0     & 0     \vspace{3pt}\\
t_2^y & 0     & 0     & 0     \vspace{3pt}\\
0     & 0     & 0     & t_3^y \vspace{3pt}\\
0     & 0     & t_2^y & 0
\end{array}\right) \nonumber\\
M&=&\left( \begin{array}{cccc}
0                & t_1^y+\gamma^y/2  & t_1^x+\gamma^x/2 & 0                 \vspace{3pt}\\
t_1^y-\gamma^y/2 & 0                 & 0                & -t_1^x-\gamma^x/2 \vspace{3pt}\\
t_1^x-\gamma^x/2 & 0                 & 0                & t_1^y+\gamma^y/2  \vspace{3pt}\\
0                & -t_1^x+\gamma^x/2 & t_1^y-\gamma^y/2 & 0
\end{array}\right), \nonumber\\
\label{eq24}
\end{eqnarray}
and ${\bm c}_{\bm r}^\dag=\left(c_{{\bm r},{\rm A}}^\dag,c_{{\bm r},{\rm B}}^\dag,c_{{\bm r},{\rm C}}^\dag,c_{{\bm r},{\rm D}}^\dag\right)$ are the creation operators of particles on sublattices A, B, C, and D, respectively. For simplicity, all the parameters are set to be real. We note that this system is in the case of $N_x=1,N_y=1$, and $q=4$ in Eq.~(\ref{eq1}). The real-space eigen-equation of this system is written in the form of Eq.~(\ref{eq17}) with ${\bm\psi}_{\bm r}=\left(\psi_{{\bm r},{\rm A}},\psi_{{\bm r},{\rm B}},\psi_{{\bm r},{\rm C}},\psi_{{\bm r},{\rm D}}\right)^{\rm T}$.

\begin{figure}[]
\includegraphics[width=7.5cm]{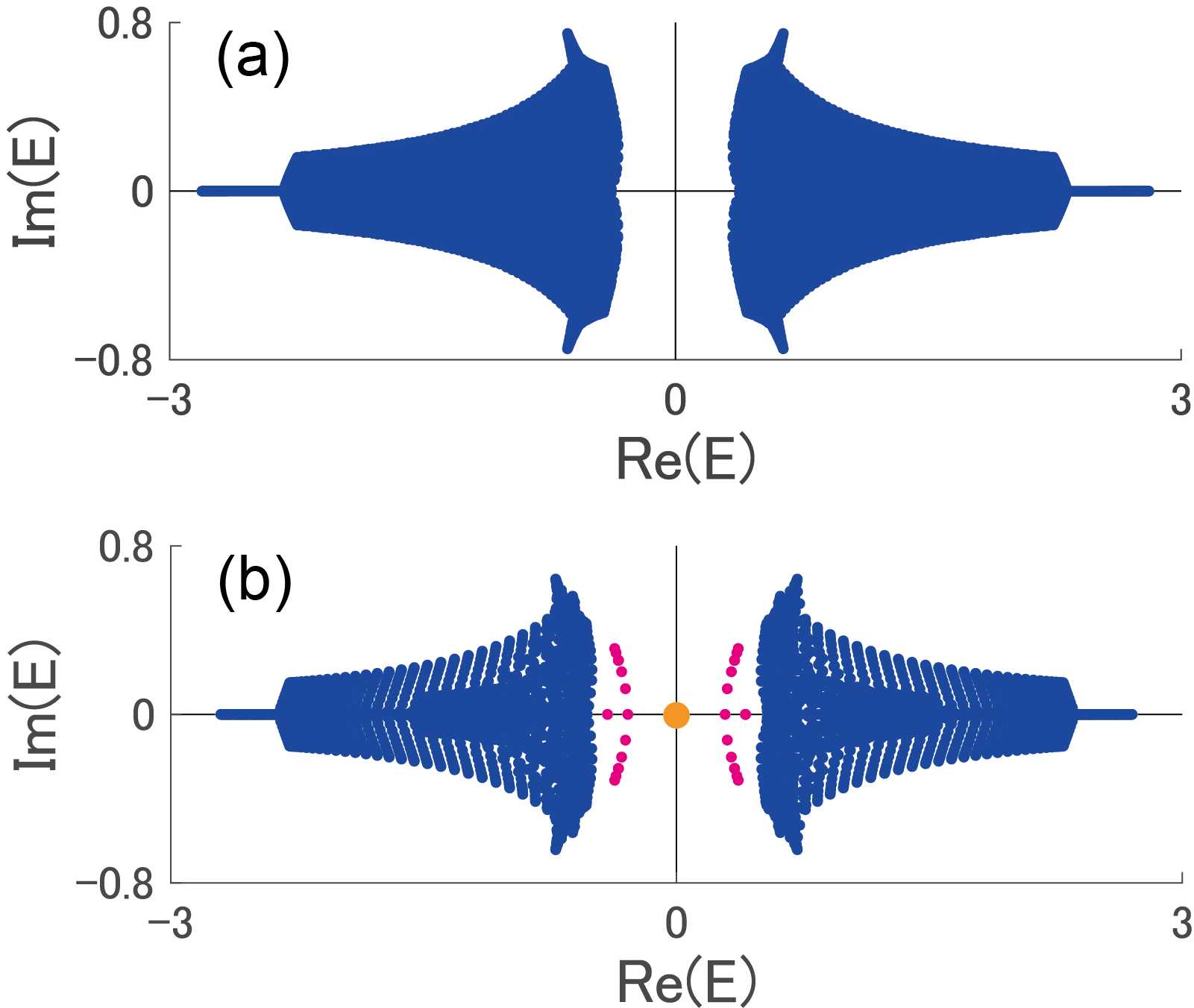}
\caption{\label{fig3}Energy eigenvalue of the non-Hermitian Benalcazar-Bernevig-Hughes model. (a) Energy band calculated from the generalized Brillouin zone. (b) Energy level in the system with $L_x=L_y=40$. In panel (b), the energy levels of the hybrid topological-skin states and the topological corner states are shown in magenta and orange, respectively. The system parameters are set to be $t_1^x=1.2,t_2^x=1,t_3^x=0.2,\gamma^x=4/3,t_1^y=0.8,t_2^y=0.9,t_3^y=0.2$, and $\gamma^y=5/3$.}
\end{figure}

In this system, we have the Bloch Hamiltonian as
\begin{equation}
H\left({\bm k}\right)=H_x\left(e^{ik_x}\right)\otimes\sigma_z+\sigma_0\otimes H_y\left(e^{ik_y}\right),
\label{eq25}
\end{equation}
where
\begin{eqnarray}
H_a\left(e^{ik_a}\right)&=&\left( \begin{array}{cc}
0                          & R_+^a\left(e^{ik_a}\right) \hspace{3pt}\\
R_-^a\left(e^{ik_a}\right) & 0
\end{array}\right), \nonumber\\
R_\pm^a\left(e^{ik_a}\right)&=&t_3^ae^{\pm ik_a}+\left(t_1^a\pm\frac{\gamma^a}{2}\right)+t_2^ae^{\mp ik_a}
\label{eq26}
\end{eqnarray}
for $a=x,y$, and $\sigma_0$ is an identity matrix, and $\sigma_z$ is the $z$ component of the Pauli matrix. We note that $H_a$ satisfies $\sigma_zH_a\sigma_z=-H_a$. Then, the characteristic equation of Eq.~(\ref{eq25}) is obtained as
\begin{equation}
\sum_{a=x,y}R_+^a\left(e^{ik_a}\right)R_-^a\left(e^{ik_a}\right)-E^2=0.
\label{eq27}
\end{equation}
Thus, we find that the system is in Case B as discussed in Sec.~\ref{sec2-2}. Therefore, from the solutions of the equations
\begin{equation}
R_+^a\left(\beta_a\right)R_-^a\left(\beta_a\right)=\lambda_a,
\label{eq28}
\end{equation}
the condition for the generalized Brillouin zone is given by
\begin{equation}
\left|\beta_{a,2}\right|=\left|\beta_{a,3}\right|,
\label{eq29}
\end{equation}
where we have
\begin{equation}
\left|\beta_{a,1}\right|\leq\dots\leq\left|\beta_{a,4}\right|
\label{eq30}
\end{equation}
for $a=x,y$. Then, we can calculate the energy bands from
\begin{equation}
E=\pm\sqrt{\lambda_x+\lambda_y}.
\label{eq31}
\end{equation}
As an example, we show the energy bands of the system obtained from the non-Bloch band theory in Fig.~\ref{fig3}(a) and confirm that they match the energy levels in a finite open plane shown in blue in Fig.~\ref{fig3}(b).

We note that in Fig.~\ref{fig3}(b), one can see that the energy levels shown in magenta do not belong to the bulk bands. These originate from the hybrid skin-topological states~\cite{Lee2019,Fu2021,Li2022,Zhu2022,Zou2021}, which result from the hybridization between topological edge states and non-Hermitian skin states. A general method of the calculation of energy bands of such states is beyond the present work. On the other hand, in some models, such as the OTY model studied in Sec.~\ref{sec3-3}, we can obtain the energy bands of the hybrid skin-topological states.

Finally, we investigate a topological phase of this system. Based on the result obtained from Ref.~\cite{Okugawa2019}, we defined the topological invariant as
\begin{equation}
\nu=w_xw_y,
\label{eq32}
\end{equation}
where
\begin{equation}
w_a=-\frac{1}{2\pi}\frac{\left[{\rm arg}R_+\left(\beta_a\right)\right]_{C_{\beta_a}}-\left[{\rm arg}R_-\left(\beta_a\right)\right]_{C_{\beta_a}}}{2}
\label{eq33}
\end{equation}
for $a=x,y$. Here, let $C_{\beta_a}$ denote a loop formed by Eq.~(\ref{eq29}) on the complex $\beta_a$-plane. In Eq.~(\ref{eq33}), $\left[{\rm arg}R_{\pm}\left(\beta_a\right)\right]_{C_{\beta_a}}$ means the change of ${\rm arg}R_{\pm}\left(\beta_a\right)$ as $\beta_a$ goes along $C_{\beta_a}$ in a counterclockwise way. In the definition, we use the fact that the non-Bloch matrix can be written in the form of Eq.~(\ref{eq25}). Importantly, the nonzero value of $\nu$ corresponds to existence of the topological corner states at $E=0$. Indeed, in the present case, the system has $w_x=1$ and $w_y=1$, which agrees with appearance of the zero-energy modes in a finite open plane as shown in Fig.~\ref{fig3}(b).

%
%

\subsection{\label{sec3-3}Okugawa-Takahashi-Yokomizo model}
\begin{figure}[]
\includegraphics[width=8.5cm]{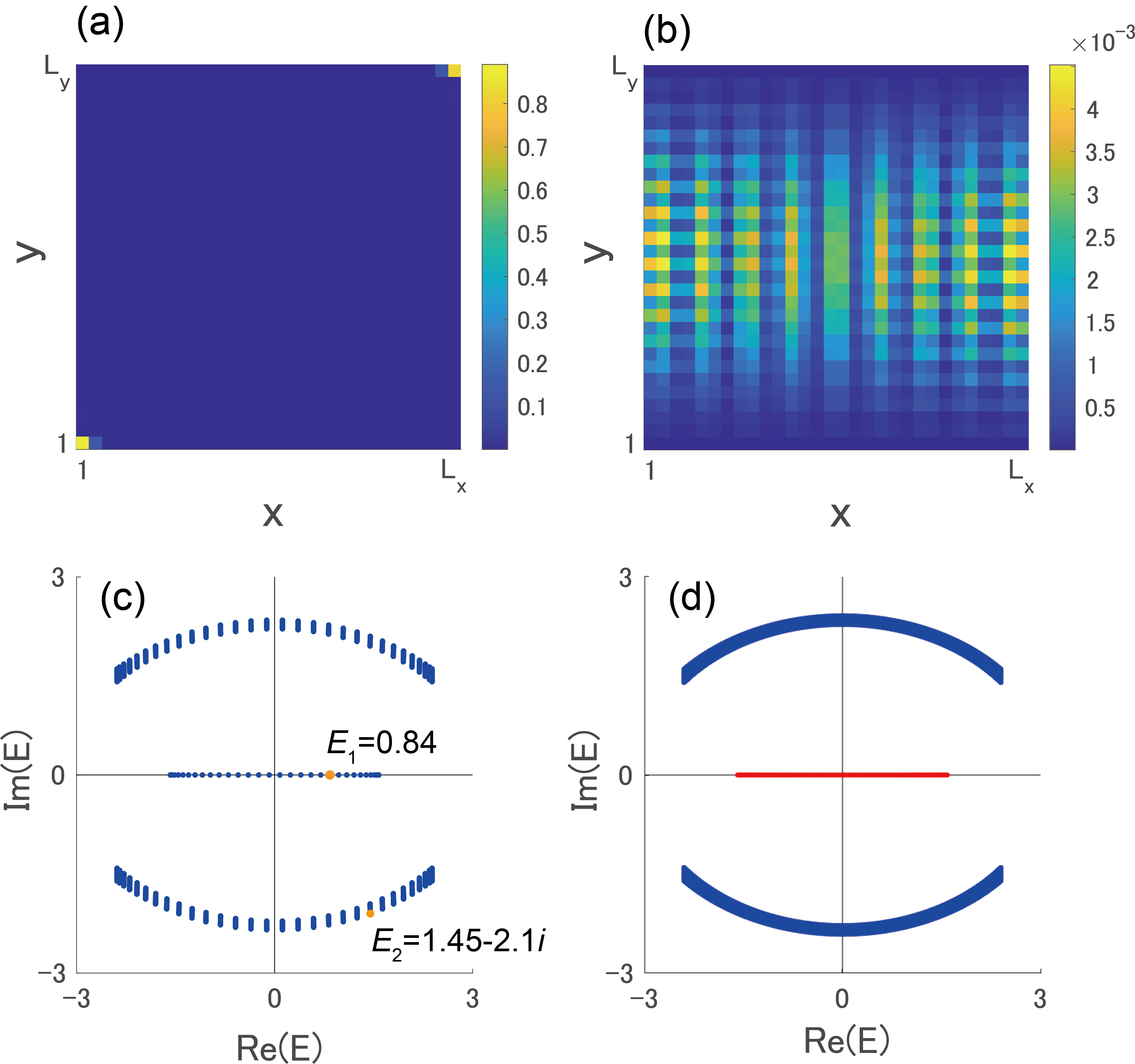}
\caption{\label{fig4}Eigenstate and energy eigenvalue in the Okugawa-Takahashi-Yokomizo model. Spatial distribution of (a) the hybrid skin-topological states and (b) the extended state in the system with $L_x=L_y=30$. The energy levels of the hybrid skin-topological states and the extended state are $E_1=0.84$ and $E_2=1.45-2.1i$, respectively. (c) Energy level in the system with $L_x=L_y=30$. (d) Energy band calculated from the non-Bloch band theory proposed in this work (blue) and that for the hybrid skin-topological states (red). We set the system parameters to be $t_x=1.2,g_x=0.9,t_y=0.8$, and $g_y=0.7$.}
\end{figure}
Finally, we study the OTY model proposed in Ref.~\cite{Okugawa2020}. The real-space Hamiltonian of this system is given by
\begin{eqnarray}
H&=&\sum_{\bm r}\left({\bm c}_{{\bm r}+\hat{\bm x}}^\dag T_x^-{\bm c}_{\bm r}+{\bm c}_{\bm n}T_x^+{\bm c}_{{\bm r}+\hat{\bm x}}\right. \nonumber\\
&&\left.+{\bm c}_{{\bm r}+\hat{\bm y}}^\dag T_y^-{\bm c}_{\bm r}+{\bm c}_{\bm r}T_y^+{\bm c}_{{\bm r}+\hat{\bm y}}\right),
\label{eq34}
\end{eqnarray}
where
\begin{eqnarray}
T_x^\pm=\left( \begin{array}{cc}
t_x\pm g_x & 0          \vspace{3pt}\\
0          & t_x\mp g_x
\end{array}\right),~T_y^\pm=\left( \begin{array}{cc}
0          & -t_y\pm g_y \vspace{3pt}\\
t_y\pm g_y & 0
\end{array}\right), \nonumber\\
\label{eq35}
\end{eqnarray}
and ${\bm c}_{\bm r}^\dag=\left(c_{{\bm r},{\rm A}}^\dag,c_{{\bm r},{\rm B}}^\dag\right)$ are the creation operators of particles on sublattices A and B, respectively. All the parameters are set to be positive real numbers. We note that this system is in the case of $N_x=1,N_y=1$, and $q=2$ in Eq.~(\ref{eq1}). The real-space eigen-equation is obtained as
\begin{equation}
T_x^-{\bm\psi}_{{\bm r}+\hat{\bm x}}+T_x^+{\bm\psi}_{{\bm r}-\hat{\bm x}}+T_y^-{\bm\psi}_{{\bm r}+\hat{\bm y}}+T_y^+{\bm\psi}_{{\bm r}-\hat{\bm y}}=E{\bm\psi}_{\bm r},
\label{eq36}
\end{equation}
where ${\bm\psi}_{\bm n}=\left(\psi_{{\bm n},{\rm A}},\psi_{{\bm n},{\rm B}}\right)^{\rm T}$. The previous work proposed that the OTY model exhibits the second-order non-Hermitian skin effect. In this effect, $O\left(L_x\right)$ (or $O\left(L_y\right)$) eigenstates are localized at the corners while $O\left(L_xL_y\right)$ eigenstates extend over the system. For example, among the energy levels of the OTY model on a finite open plane with $L_x=L_y=30$ as shown in Fig.~\ref{fig4}(c), the eigenstates with $E_1=0.84$ are localized at the corners [Fig.~\ref{fig4}(a)] while the eigenstate with $E_2=1.45-2.1i$ extends over the system [Fig.~\ref{fig4}(b)]. We note that the second-order non-Hermitian skin state in this model is a kind of the hybrid skin-topological state~\cite{Fu2021}.

Now, we calculate the energy bands of the OTY model from the non-Bloch band theory. In this system, we have the characteristic equation of the Bloch Hamiltonian as
\begin{eqnarray}
R_x\left(e^{ik_x},E\right)+R_y\left(e^{ik_y}\right)=0,
\label{eq37}
\end{eqnarray}
where
\begin{eqnarray}
R_x\left(e^{ik_x},E\right)&=&\left[t_x\left(e^{ik_x}+e^{-ik_x}\right)-E\right]^2 \nonumber\\
&&-g_x^2\left(e^{ik_x}-e^{-ik_x}\right)^2, \nonumber\\
R_y\left(e^{ik_y}\right)&=&t_y^2\left(e^{ik_y}+e^{-ik_y}\right)^2-g_y^2\left(e^{ik_y}-e^{-ik_y}\right)^2. \nonumber\\
\label{eq38}
\end{eqnarray}
Hence, this system is in Case B as discussed in Sec.~\ref{sec2-2}. Then, the conditions for the generalized Brillouin zone are given by the solutions of the quadratic equation $R_x\left(\beta_x,E\right)=\lambda$ for $\beta_x$ and those of the quadratic equation $R_y\left(\beta_y\right)=-\lambda$ for $\beta_y$. In fact, the condition is obtained as
\begin{equation}
\left|\beta_{a,2}\right|=\left|\beta_{a,3}\right|,~\left(a=x,y\right),
\label{eq39}
\end{equation}
where the solutions satisfy Eq.~(\ref{eq3}). At last, we can calculate the energy bands by combining the generalized Brillouin zone and Eq.~(\ref{eq37}). Importantly, the generalized Brillouin zone spanned by $\left(\beta_x,\beta_y\right)~\left(\beta_x\equiv e^{ik_x},\beta_y\equiv e^{ik_y}\right)$ becomes a torus defined by $\left|\beta_x\right|=1$ and $\left|\beta_y\right|=1$ in this case because the Bloch wavevector $\left(k_x,k_y\right)$ takes real values. Then, we obtain the energy bands as
\begin{eqnarray}
E\left(k_x,k_y\right)&=&2t_x\cos k_x \nonumber\\
&&\pm2i\sqrt{g_x^2\sin^2k_x+t_y^2\cos^2k_y+g_y^2\sin^2k_y}. \nonumber\\
\label{eq40}
\end{eqnarray}
The results are shown in blue in Fig.~\ref{fig4}(d). Comparing Fig.~\ref{fig4}(c) with Fig.~\ref{fig4}(d), we see that the blue energy bands reproduces the bulk energy levels in a finite open plane. Meanwhile, the energy levels of the hybrid skin-topological states are not obtained from the non-Bloch band theory proposed in this work.

Next, we explain a way to calculate the energy band of the hybrid skin-topological state. In the following, we set $t_y>g_y$ for convenience. The key ingredient is that $\beta_y$ can be determined as a constant so that the non-Bloch matrix
\begin{equation}
{\cal H}\left(\beta_x,\beta_y\right)=\beta_xT_x^-+\beta_x^{-1}T_x^++\beta_yT_y^-+\beta_y^{-1}T_y^+
\label{eq41}
\end{equation}
has eigenvectors independent of the value of $\beta_x$, such as $\left(1,0\right)^{\rm T}$ and $\left(0,1\right)^{\rm T}$. In this case, a non-Bloch wave for $\beta_x$ can be constructed. In fact, we can take an ansatz
\begin{eqnarray}
\left\{ \begin{array}{l}
\displaystyle{\bm\psi}_{\bm r}=\Psi_{n_x}{\bm\Phi}_{n_y}, \vspace{5pt}\\
\displaystyle\Psi_{n_x}=\sum_j\left(\beta_{x,j}\right)^{n_x}\upsilon^{\left(j\right)}, \vspace{5pt}\\
\displaystyle{\bm\Phi}_{n_y}=\sum_j\left(\beta_{y,j}\right)^{n_y}\chi^{\left(j\right)}\left( \begin{array}{c}
0 \\
1
\end{array}\right)
\end{array}\right.
\label{eq42}
\end{eqnarray}
for Eq.~(\ref{eq36}), where $\upsilon^{\left(j\right)}$ and $\chi^{\left(j\right)}$ are constants. By imposing $\left(0,1\right)^{\rm T}$ to be an eigenvector of ${\cal H}\left(\beta_x,\beta_y\right)$, the $\left(1,2\right)$ component of ${\cal H}\left(\beta_x,\beta_y\right)$ should vanish, and possible values of $\beta_y$ are given by
\begin{equation}
\beta_{y,1}=-i\sqrt{\frac{t_y-g_y}{t_y+g_y}},~\beta_{y,2}=i\sqrt{\frac{t_y-g_y}{t_y+g_y}}.
\label{eq43}
\end{equation}
In this case, by putting $\chi^{\left(1\right)}=-\chi^{\left(2\right)}$, the wave function satisfies an open boundary condition along the $y$ direction
\begin{equation}
{\bm\Phi}_0={\bm0}.
\label{eq44}
\end{equation}
We note that the wave function is located near $n_y=1$ because $t_y>g_y>0$. Then, the energy eigenvalue of ${\cal H}\left(\beta_x,\beta_y\right)$ is given by
\begin{equation}
\left(t_x+g_x\right)\beta_x+\frac{t_x-g_x}{\beta_x}=E,
\label{eq45}
\end{equation}
Therefore, the condition for the generalized Brillouin zone for the hybrid skin-topological state is obtained from the two solutions of Eq.~(\ref{eq45}) as
\begin{equation}
\left|\beta_{x,1}\right|=\left|\beta_{x,2}\right|.
\label{eq46}
\end{equation}
Finally, we can get the energy band as
\begin{equation}
E=2\sqrt{t_x^2-g_x^2}\cos\theta,
\label{eq47}
\end{equation}
where $\theta$ is real. We note that the localization lengths of the hybrid skin-topological state in the $x$ and $y$ directions can be explicitly written as
\begin{equation}
\lambda_x=\frac{1}{\log\sqrt{\left(t_x-g_x\right)/\left(t_x+g_x\right)}}
\label{eq48}
\end{equation}
and
\begin{equation}
\lambda_y=\frac{1}{\log\sqrt{\left(t_y-g_y\right)/\left(t_y+g_y\right)}},
\label{eq49}
\end{equation}
respectively.

Importantly, we can take another ansatz of the hybrid skin-topological state
\begin{eqnarray}
\left\{ \begin{array}{l}
\displaystyle{\bm\psi}_{\bm r}=\Psi_{n_x}{\bm\Phi}_{n_y}, \vspace{5pt}\\
\displaystyle\Psi_{n_x}=\sum_j\left(\beta_{x,j}\right)^{n_x}\bar{\upsilon}^{\left(j\right)}, \vspace{5pt}\\
\displaystyle{\bm\Phi}_{n_y}=\sum_j\left(\beta_{y,j}\right)^{n_y-L_y-1}\bar{\chi}^{\left(j\right)}\left( \begin{array}{c}
1 \\
0
\end{array}\right).
\end{array}\right.
\label{eq50}
\end{eqnarray}
Here, $\bar{\upsilon}^{\left(j\right)}$ and $\bar{\chi}^{\left(j\right)}$ are constants. By imposing $\left(1,0\right)^{\rm T}$ to be an eigenvector of ${\cal H}\left(\beta_x,\beta_y\right)$, the allowed values of $\beta_y$ are given by
\begin{equation}
\beta_{y,1}=-i\sqrt{\frac{t_y+g_y}{t_y-g_y}},~\beta_{y,2}=i\sqrt{\frac{t_y+g_y}{t_y-g_y}},
\label{eq51}
\end{equation}
In this case, by taking $\bar{\chi}^{\left(1\right)}=-\bar{\chi}^{\left(2\right)}$, the wave function satisfies an open boundary condition along the $y$ direction
\begin{equation}
{\bm\Phi}_{L_y+1}={\bm0},
\label{eq52}
\end{equation}
and it is located near $n_y=L_y$. Finally, the energy eigenvalue of ${\cal H}\left(\beta_x,\beta_y\right)$ is given by
\begin{equation}
\left(t_x-g_x\right)\beta_x+\frac{t_x+g_x}{\beta_x}=E,
\label{eq53}
\end{equation}
and we can get the energy band of this hybrid skin-topological state in the form of Eq.~(\ref{eq47}). We note that the localization lengths of the hybrid skin-topological state in the $x$ and $y$ directions can be explicitly written as
\begin{equation}
\lambda_x=\frac{1}{\log\sqrt{\left(t_x+g_x\right)/\left(t_x-g_x\right)}}
\label{eq54}
\end{equation}
and
\begin{equation}
\lambda_y=\frac{1}{\log\sqrt{\left(t_y+g_y\right)/\left(t_y-g_y\right)}},
\label{eq55}
\end{equation}
respectively.

The energy band for the hybrid skin-topological state obtained above is shown in red in Fig.~\ref{fig4}(d). Comparing Fig.~\ref{fig4}(c) with Fig.~\ref{fig4}(d), we confirm that our calculation reproduces the energy levels of the hybrid skin-topological states in a finite open plain.

%
%

\section{\label{sec4}Discussion}
\begin{figure}[]
\includegraphics[width=8.5cm]{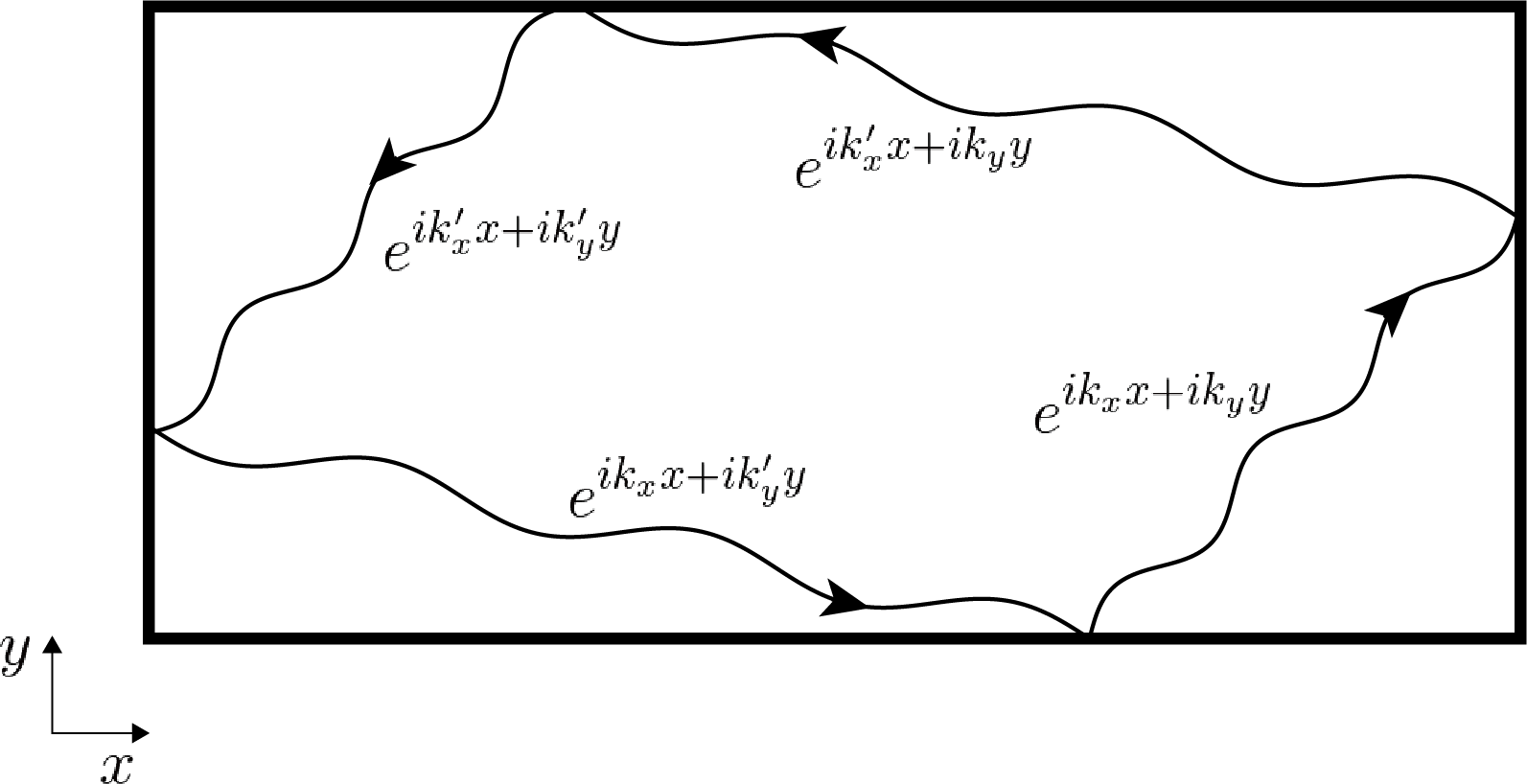}
\caption{\label{fig5}Reflection of the plane waves at the boundaries in the two-dimensional non-Hermitian system. Through the reflection at the boundary parallel to the $y$ direction, the wavenumber $k_x$ changes to $k_x^\prime$, but the wavenumber $k_y$ is unchanged. The situation is similar at the boundary parallel to the $x$ direction.}
\end{figure}
In this section, we explain the reason why our non-Bloch band theory is established in the 2D non-Hermitian system. The key ingredient of our work is that the ansatz of the real-space eigen-equation can be written as Eqs.~(\ref{eq5}) and (\ref{eq11}) in Case A and Case B, respectively. This is related to the condition for the standing-wave formation. In the bulk, the standing wave is formed by interference between plane waves which are generated by reflection at the boundaries of the system. In general, the process of the standing-wave formation is complex. In some cases, the condition for the standing-wave formation becomes straightforward. In Fig.~\ref{fig5}, we show such a case in the rectangle geometry, where the standing wave consists of four plane waves. The plane wave $e^{ik_xx+ik_yy}$ preserves the wavenumber in the direction parallel to the boundary where the plane wave is reflected. In the present work, we propose the two cases in which the above situation can be realized. In Case A, the reflection of the plane wave becomes simple because of the suppression of the non-Hermitian skin effect in one direction. In Case B, since the motions of the plane wave in the $x$ direction and the $y$ direction are decoupled, we can regard the 2D system as two 1D systems. Finally, the leading terms of the standing wave have the form of the linear combination of a few plane waves in both cases.

\begin{figure}[]
\includegraphics[width=8.5cm]{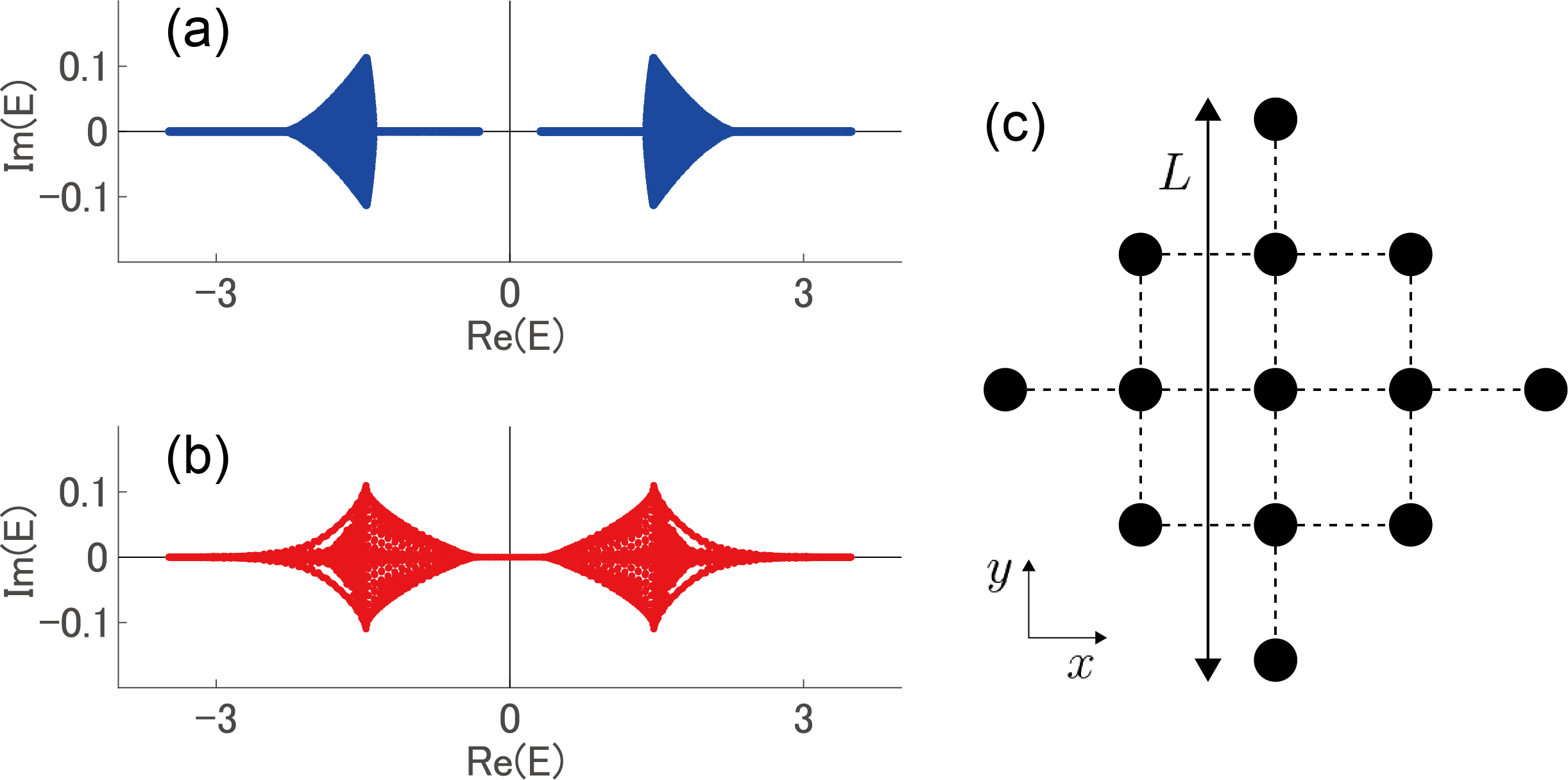}
\caption{\label{fig6}Geometry dependence of the energy bands in the model of the non-Hermitian Chern insulator (\ref{eq5}). (a) Energy band calculated from the non-Bloch band theory for the rectangle geometry proposed in this work. (b) Energy level in a finite open plane on the diamond geometry with $L=59$. (c) Diamond geometry. In panels (a) and (b), we set the system parameters same as Figs.~\ref{fig2}(a)-(d).}
\end{figure}

Based on the above concept, it naturally follows that the geometry of the system largely affects the condition for the standing-wave formation. In fact, the non-Hermitian skin effect in a 2D system depends on geometry~\cite{Zhang2022} Namely, the non-Bloch band theory proposed in this work should depend crucially on geometries. For example, in the model of the non-Hermitian Chern insulator as discussed in Sec.~\ref{sec3-1}, the energy bands obtained from the non-Bloch band theory for the rectangle geometry [Fig.~\ref{fig6}(a), which is the same as Fig.~\ref{fig2}(c)] do not match the energy levels in a finite open plane [Fig.~\ref{fig6}(b)] on the diamond geometry [Fig.~\ref{fig6}(c)]. This indicates that non-Bloch bands on the rectangle geometry are different from those on the diamond geometry. Thus, we conclude that the energy bands in a 2D non-Hermitian system depend on geometries. Construction of a general method to calculate the energy bands in a 2D non-Hermitian system with various geometries is left for a future work.

Finally, we focus on the model of the non-Hermitian Chern insulator with $\gamma_x\neq0$ and $\gamma_y\neq0$. We note that the non-Hermitian skin effect occurs in both the $x$ and the $y$ directions, and the system is not in Case A. In this case, the plane wave is scattered at the boundary into numerous waves. Since the reflection process of the plane wave becomes complex compared to the case in Fig.~\ref{fig5}, we cannot find the condition for the standing-wave formation, and the standing wave cannot be written in the form of Eq.~(\ref{eq5}). Hence, it is unclear whether we can construct the non-Bloch band theory in the model of the non-Hermitian Chern insulator with arbitrary parameters, and also in general 2D non-Hermitian systems.

%
%

\section{\label{sec5}Summary}
In this paper, we propose two classes of the 2D non-Hermitian tight-binding systems in which the non-Bloch band theory can be constructed. In terms of the present non-Bloch band theory, we calculate the generalized Brillouin zone and the energy bands in the model of the non-Hermitian Chern insulator, the non-Hermitian BBH model, and the OTY model. We confirm that the energy bands reproduce the bulk energy levels in a finite open plane on the rectangle geometry. Furthermore, we investigate topological properties of theses models from the non-Bloch band theory. In the model of the non-Hermitian Chern insulator and the non-Hermitian BBH model, it is shown that the topological invariant defined from the generalized Brillouin zone can predict existence of the topological edge states. In the OTY model, we propose the method to calculate the energy bands for the hybrid skin-topological state. We note that the construction of the non-Bloch band theory in general 2D non-Hermitian system and a general method to calculate the energy bands for the hybrid skin-topological state are left for a future work.

%
%

\begin{acknowledgements}
We are grateful to Ryo Okugawa for variable discussion. This work was supported by JSPS KAKENHI Grant No.~22H00108 and No.~JP21J01409.
\end{acknowledgements}

%
%

\appendix

%
%

\section{\label{secA}Non-Bloch band theory in the case with $q=1$}
In this appendix, we construct the non-Bloch band theory in the system described by the Hamiltonian (\ref{eq1}) with $q=1$. For example, the non-Hermitian skin effect in this case was investigated in a synthetic photonic lattice~\cite{YSong2020}. When $q=1$, the real-space eigen-equation (\ref{eq2}) is explicitly written as
\begin{equation}
\sum_{i=-N_x}^{N_x}t_i^{\left(x\right)}\psi_{{\bm r}+i\hat{\bm x}}+\sum_{i=-N_y}^{N_y}t_i^{\left(y\right)}\psi_{{\bm r}+i\hat{\bm y}}=E\psi_{\bm r},
\label{eqapp1a}
\end{equation}
where ${\bm r}=\left(n_x,n_y\right)$. When we take
\begin{equation}
\psi_{\bm r}=\varphi_{n_x}^{\left(x\right)}\varphi_{n_y}^{\left(y\right)}
\label{eqapp2a}
\end{equation}
as an ansatz for Eq.~(\ref{eqapp1a}), we get
\begin{eqnarray}
\sum_{i=-N_a}^{N_a}t_i^{\left(a\right)}\varphi_{n_a+i}^{\left(a\right)}&=&\lambda_a\varphi_{n_a}^{\left(a\right)},~\left(a=x,y\right), \label{eqapp3a}\\
\lambda_x+\lambda_y&=&E. \label{eqapp4a}
\end{eqnarray}
Here, the solutions of Eq.~(\ref{eqapp3a}) can be given in the form of a linear combination. Namely, we obtain
\begin{equation}
\varphi_{n_a}^{\left(a\right)}=\sum_j\left(\beta_{a,j}\right)^{n_a}\phi_j,
\label{eqapp5a}
\end{equation}
where $\beta_a=\beta_{a,j}$ is the solution of the characteristic equation
\begin{equation}
\sum_{m=-N_a}^{N_a}t_i^{\left(a\right)}\left(\beta_a\right)^m-\lambda_a=0
\label{eqapp6a}
\end{equation}
for $a=x,y$. By following the conventional non-Bloch band theory, we can obtain the condition for the generalized Brillouin zone in terms of the solutions of Eq.~(\ref{eqapp6a}), given by
\begin{equation}
\left|\beta_{a,N_a}\right|=\left|\beta_{a,N_a+1}\right|,
\label{eqapp7a}
\end{equation}
where
\begin{equation}
\left|\beta_{a,i}\right|\leq\dots\leq\left|\beta_{a,2N_a}\right|
\label{eqapp8a}
\end{equation}
for $a=x,y$. Thus, a set of the trajectories of $\beta_{a,N_a}$ and $\beta_{a,N_a+1}$ satisfying Eq.~(\ref{eqapp7a}) forms the generalized Brillouin zone. Finally, from $\beta_x$ and $\beta_y$ on the generalized Brillouin zone, we can calculate the energy band by using Eq.~(\ref{eqapp4a}).

%
%

\section{\label{secB}Quantization of the Chern number}
\begin{figure}[]
\includegraphics[width=6cm]{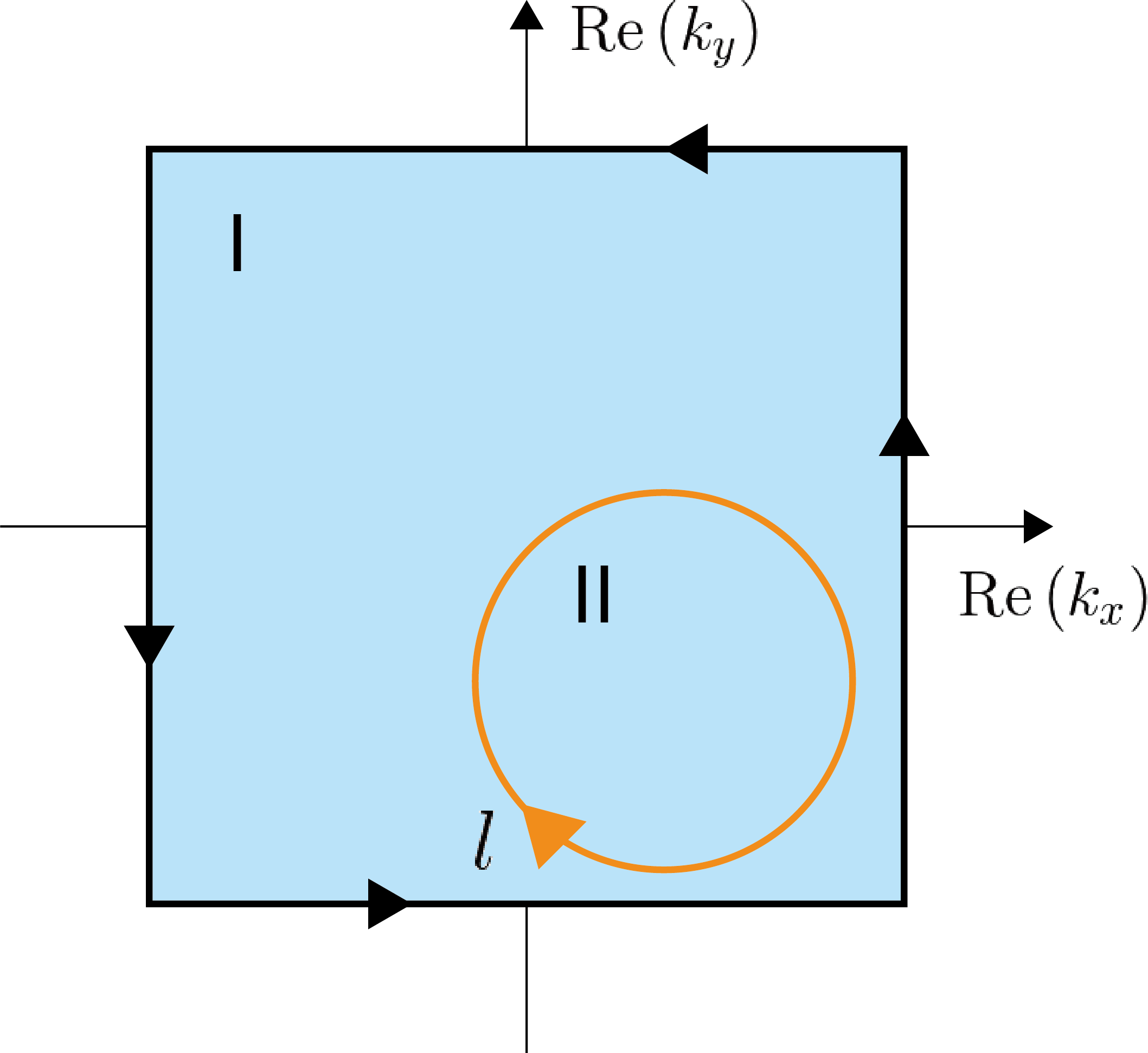}
\caption{\label{fig7}Generalized Brillouin zone on the $k_x\mathchar`-k_y$ plane for the complex Bloch wavevector ${\bm k}$. The imaginary part of ${\bm k}$ is omitted. $l$ represents the boundary between region I and region I\hspace{-.1em}I.}
\end{figure}
In this appendix, we show that the values of the Chern number defined in Eq.~(\ref{eq20}) are quantized when the band $n$ is separated from other bands by a line gap. We assume that the system is in Case A or Case B, and the generalized Brillouin zone can be constructed. For the non-Bloch matrix obtained from Eq.~(\ref{eq1}), let $\left|u_{R,n}\left({\bm k}\right)\right\rangle$ and $\left\langle u_{L,n}\left({\bm k}\right)\right|$ denote the right and left eigenvectors, respectively. Here, ${\bm k}$ is the complex Bloch wavevector, and $n$ expresses the band index. We impose that the right and left eigenvectors satisfy
\begin{equation}
\left\langle u_{L,m}\left({\bm k}\right)\middle|u_{R,n}\left({\bm k}\right)\right\rangle=\delta_{m,n}.
\label{eqapp1b}
\end{equation}
Furthermore, we define the Berry connection as
\begin{equation}
\mathcal{\bm A}_n\left({\bm k}\right)=\left\langle u_{L,n}\left({\bm k}\right)\middle|{\bm\nabla}_{\bm k}\middle|u_{R,n}\left({\bm k}\right)\right\rangle.
\label{eqapp2b}
\end{equation}

First of all, we discuss the generalized Brillouin zone in this work. In our 2D non-Hermitian system, let us regard $k_y$ as a system parameter. Under a given value of $k_y$, the system is regarded as a 1D non-Hermitian system described by the complex Bloch wavenumber $k_x$. Then, a complex parameter $e^{ik_x}$ forms a loop on the complex plane~\cite{Yang2020}. This leads to numerous loops of $e^{ik_x}$ for all values of $k_y$. Similarly, $e^{ik_y},~k_y\in{\mathbb C}$ forms a loop for all values of $k_x$. Therefore, the whole generalized Brillouin zone in the 2D non-Hermitian system forms a closed surface. For example, the generalized Brillouin zone in the Chern number discussed in Sec.~\ref{sec2-1} becomes a closed surface [Fig.~\ref{fig2}(b)].

Topologically nontrivial phases can be associated with topological obstruction for a single gauge choice over the whole ${\bm k}$ space. Therefore, in the present case, we separate the generalized Brillouin zone into two regions so that we take gauge I on one region and gauge I\hspace{-.1em}I on the other region [Fig.~\ref{fig7}]. Importantly, the two gauges are related by the gauge transformation
\begin{eqnarray}
\left|\psi_{R,n}^{\rm I\hspace{-.1em}I}\left({\bm k}\right)\right\rangle&=&r_n\left({\bm k}\right)e^{i\chi_n\left({\bm k}\right)}\left|\psi_{R,n}^{\rm I}\left({\bm k}\right)\right\rangle, \label{eqapp3b}\\
\left\langle\psi_{L,n}^{\rm I\hspace{-.1em}I}\left({\bm k}\right)\right|&=&\frac{1}{r_n\left({\bm k}\right)}e^{-i\chi_n\left({\bm k}\right)}\left\langle\psi_{L,n}^{\rm I}\left({\bm k}\right)\right|, \label{eqapp4b}
\end{eqnarray}
where $r_n\left({\bm k}\right)$ and $\chi_n\left({\bm k}\right)$ are real functions~\cite{Shen2018}. Here, from the Stokes' theorem, Eq.~(\ref{eq20}) can be rewritten in terms of the corresponding Berry connection as
\begin{equation}
C_n=\frac{1}{2\pi i}\oint_l d{\bm k}\cdot\left(\mathcal{\bm A}^{\rm I}_n\left({\bm k}\right)-\mathcal{\bm A}^{\rm I\hspace{-.1em}I}_n\left({\bm k}\right)\right),
\label{eqapp5b}
\end{equation}
where $l$ denote the boundary between region I and region I\hspace{-.1em}I. Hence, we obtain
\begin{equation}
C_n=\frac{1}{2\pi}\oint_ld{\bm k}\cdot{\bm\nabla}_{\bm k}\chi_n\left({\bm k}\right),
\label{eqapp6b}
\end{equation}
where we use
\begin{equation}
\mathcal{\bm A}^{\rm I\hspace{-.1em}I}_n\left({\bm k}\right)=\mathcal{\bm A}^{\rm I}_n\left({\bm k}\right)+i{\bm\nabla}_{\bm k}\chi_n\left({\bm k}\right)+\frac{{\bm\nabla}_{\bm k}r_n\left({\bm k}\right)}{r_n\left({\bm k}\right)}
\label{eqapp7b}
\end{equation}
and
\begin{equation}
\oint_ld{\bm k}\cdot{\bm\nabla}_{\bm k}\log r_n\left({\bm k}\right)=0.
\label{eqapp8b}
\end{equation}
The integral in Eq.~(\ref{eqapp6b}) is always equal to an integer multiple of $2\pi$ because the wave functions are single-valued. As a result, the Chern number defined in Eq.~(\ref{eq20}) takes quantized values when the band $n$ is separated from other bands by a line gap.

%
%

%

\providecommand{\noopsort}[1]{}\providecommand{\singleletter}[1]{#1}%

\begin{thebibliography}{71}%
\makeatletter
\providecommand \@ifxundefined [1]{%
 \@ifx{#1\undefined}
}%
\providecommand \@ifnum [1]{%
 \ifnum #1\expandafter \@firstoftwo
 \else \expandafter \@secondoftwo
 \fi
}%
\providecommand \@ifx [1]{%
 \ifx #1\expandafter \@firstoftwo
 \else \expandafter \@secondoftwo
 \fi
}%
\providecommand \natexlab [1]{#1}%
\providecommand \enquote  [1]{``#1''}%
\providecommand \bibnamefont  [1]{#1}%
\providecommand \bibfnamefont [1]{#1}%
\providecommand \citenamefont [1]{#1}%
\providecommand \href@noop [0]{\@secondoftwo}%
\providecommand \href [0]{\begingroup \@sanitize@url \@href}%
\providecommand \@href[1]{\@@startlink{#1}\@@href}%
\providecommand \@@href[1]{\endgroup#1\@@endlink}%
\providecommand \@sanitize@url [0]{\catcode `\\12\catcode `\$12\catcode
  `\&12\catcode `\#12\catcode `\^12\catcode `\_12\catcode `\%12\relax}%
\providecommand \@@startlink[1]{}%
\providecommand \@@endlink[0]{}%
\providecommand \url  [0]{\begingroup\@sanitize@url \@url }%
\providecommand \@url [1]{\endgroup\@href {#1}{\urlprefix }}%
\providecommand \urlprefix  [0]{URL }%
\providecommand \Eprint [0]{\href }%
\providecommand \doibase [0]{https://doi.org/}%
\providecommand \selectlanguage [0]{\@gobble}%
\providecommand \bibinfo  [0]{\@secondoftwo}%
\providecommand \bibfield  [0]{\@secondoftwo}%
\providecommand \translation [1]{[#1]}%
\providecommand \BibitemOpen [0]{}%
\providecommand \bibitemStop [0]{}%
\providecommand \bibitemNoStop [0]{.\EOS\space}%
\providecommand \EOS [0]{\spacefactor3000\relax}%
\providecommand \BibitemShut  [1]{\csname bibitem#1\endcsname}%
\let\auto@bib@innerbib\@empty
\bibitem [{\citenamefont {Ashida}\ \emph {et~al.}(2020)\citenamefont {Ashida},
  \citenamefont {Gong},\ and\ \citenamefont {Ueda}}]{Ashida2020}%
  \BibitemOpen
  \bibfield  {author} {\bibinfo {author} {\bibfnamefont {Y.}~\bibnamefont
  {Ashida}}, \bibinfo {author} {\bibfnamefont {Z.}~\bibnamefont {Gong}},\ and\
  \bibinfo {author} {\bibfnamefont {M.}~\bibnamefont {Ueda}},\ }\href@noop {}
  {\bibfield  {journal} {\bibinfo  {journal} {Adv. Phys.}\ }\textbf {\bibinfo
  {volume} {69}},\ \bibinfo {pages} {249} (\bibinfo {year} {2020})}\BibitemShut
  {NoStop}%
\bibitem [{\citenamefont {Yao}\ and\ \citenamefont {Wang}(2018)}]{Yao2018}%
  \BibitemOpen
  \bibfield  {author} {\bibinfo {author} {\bibfnamefont {S.}~\bibnamefont
  {Yao}}\ and\ \bibinfo {author} {\bibfnamefont {Z.}~\bibnamefont {Wang}},\
  }\href {https://doi.org/10.1103/PhysRevLett.121.086803} {\bibfield  {journal}
  {\bibinfo  {journal} {Phys. Rev. Lett.}\ }\textbf {\bibinfo {volume} {121}},\
  \bibinfo {pages} {086803} (\bibinfo {year} {2018})}\BibitemShut {NoStop}%
\bibitem [{\citenamefont {Longhi}(2019)}]{Longhi2019}%
  \BibitemOpen
  \bibfield  {author} {\bibinfo {author} {\bibfnamefont {S.}~\bibnamefont
  {Longhi}},\ }\href {https://doi.org/10.1103/PhysRevResearch.1.023013}
  {\bibfield  {journal} {\bibinfo  {journal} {Phys. Rev. Research}\ }\textbf
  {\bibinfo {volume} {1}},\ \bibinfo {pages} {023013} (\bibinfo {year}
  {2019})}\BibitemShut {NoStop}%
\bibitem [{\citenamefont {Song}\ \emph {et~al.}(2019)\citenamefont {Song},
  \citenamefont {Yao},\ and\ \citenamefont {Wang}}]{Song2019}%
  \BibitemOpen
  \bibfield  {author} {\bibinfo {author} {\bibfnamefont {F.}~\bibnamefont
  {Song}}, \bibinfo {author} {\bibfnamefont {S.}~\bibnamefont {Yao}},\ and\
  \bibinfo {author} {\bibfnamefont {Z.}~\bibnamefont {Wang}},\ }\href
  {https://doi.org/10.1103/PhysRevLett.123.170401} {\bibfield  {journal}
  {\bibinfo  {journal} {Phys. Rev. Lett.}\ }\textbf {\bibinfo {volume} {123}},\
  \bibinfo {pages} {170401} (\bibinfo {year} {2019})}\BibitemShut {NoStop}%
\bibitem [{\citenamefont {Okuma}\ \emph {et~al.}(2020)\citenamefont {Okuma},
  \citenamefont {Kawabata}, \citenamefont {Shiozaki},\ and\ \citenamefont
  {Sato}}]{Okuma2020}%
  \BibitemOpen
  \bibfield  {author} {\bibinfo {author} {\bibfnamefont {N.}~\bibnamefont
  {Okuma}}, \bibinfo {author} {\bibfnamefont {K.}~\bibnamefont {Kawabata}},
  \bibinfo {author} {\bibfnamefont {K.}~\bibnamefont {Shiozaki}},\ and\
  \bibinfo {author} {\bibfnamefont {M.}~\bibnamefont {Sato}},\ }\href
  {https://doi.org/10.1103/PhysRevLett.124.086801} {\bibfield  {journal}
  {\bibinfo  {journal} {Phys. Rev. Lett.}\ }\textbf {\bibinfo {volume} {124}},\
  \bibinfo {pages} {086801} (\bibinfo {year} {2020})}\BibitemShut {NoStop}%
\bibitem [{\citenamefont {Zhang}\ \emph {et~al.}(2020)\citenamefont {Zhang},
  \citenamefont {Yang},\ and\ \citenamefont {Fang}}]{Zhang2020}%
  \BibitemOpen
  \bibfield  {author} {\bibinfo {author} {\bibfnamefont {K.}~\bibnamefont
  {Zhang}}, \bibinfo {author} {\bibfnamefont {Z.}~\bibnamefont {Yang}},\ and\
  \bibinfo {author} {\bibfnamefont {C.}~\bibnamefont {Fang}},\ }\href
  {https://doi.org/10.1103/PhysRevLett.125.126402} {\bibfield  {journal}
  {\bibinfo  {journal} {Phys. Rev. Lett.}\ }\textbf {\bibinfo {volume} {125}},\
  \bibinfo {pages} {126402} (\bibinfo {year} {2020})}\BibitemShut {NoStop}%
\bibitem [{\citenamefont {McDonald}\ and\ \citenamefont
  {Clerk}(2020)}]{Mcdonald2020}%
  \BibitemOpen
  \bibfield  {author} {\bibinfo {author} {\bibfnamefont {A.}~\bibnamefont
  {McDonald}}\ and\ \bibinfo {author} {\bibfnamefont {A.~A.}\ \bibnamefont
  {Clerk}},\ }\href@noop {} {\bibfield  {journal} {\bibinfo  {journal} {Nat.
  Commun.}\ }\textbf {\bibinfo {volume} {11}},\ \bibinfo {pages} {5382}
  (\bibinfo {year} {2020})}\BibitemShut {NoStop}%
\bibitem [{\citenamefont {Yi}\ and\ \citenamefont {Yang}(2020)}]{Yi2020}%
  \BibitemOpen
  \bibfield  {author} {\bibinfo {author} {\bibfnamefont {Y.}~\bibnamefont
  {Yi}}\ and\ \bibinfo {author} {\bibfnamefont {Z.}~\bibnamefont {Yang}},\
  }\href {https://doi.org/10.1103/PhysRevLett.125.186802} {\bibfield  {journal}
  {\bibinfo  {journal} {Phys. Rev. Lett.}\ }\textbf {\bibinfo {volume} {125}},\
  \bibinfo {pages} {186802} (\bibinfo {year} {2020})}\BibitemShut {NoStop}%
\bibitem [{\citenamefont {Li}\ \emph {et~al.}(2020)\citenamefont {Li},
  \citenamefont {Lee}, \citenamefont {Mu},\ and\ \citenamefont
  {Gong}}]{Li2020}%
  \BibitemOpen
  \bibfield  {author} {\bibinfo {author} {\bibfnamefont {L.}~\bibnamefont
  {Li}}, \bibinfo {author} {\bibfnamefont {C.~H.}\ \bibnamefont {Lee}},
  \bibinfo {author} {\bibfnamefont {S.}~\bibnamefont {Mu}},\ and\ \bibinfo
  {author} {\bibfnamefont {J.}~\bibnamefont {Gong}},\ }\href@noop {} {\bibfield
   {journal} {\bibinfo  {journal} {Nat. Commun.}\ }\textbf {\bibinfo {volume}
  {11}},\ \bibinfo {pages} {5491} (\bibinfo {year} {2020})}\BibitemShut
  {NoStop}%
\bibitem [{\citenamefont {Yokomizo}\ and\ \citenamefont
  {Murakami}(2021{\natexlab{a}})}]{Yokomizo2021v2}%
  \BibitemOpen
  \bibfield  {author} {\bibinfo {author} {\bibfnamefont {K.}~\bibnamefont
  {Yokomizo}}\ and\ \bibinfo {author} {\bibfnamefont {S.}~\bibnamefont
  {Murakami}},\ }\href {https://doi.org/10.1103/PhysRevB.104.165117} {\bibfield
   {journal} {\bibinfo  {journal} {Phys. Rev. B}\ }\textbf {\bibinfo {volume}
  {104}},\ \bibinfo {pages} {165117} (\bibinfo {year}
  {2021}{\natexlab{a}})}\BibitemShut {NoStop}%
\bibitem [{\citenamefont {Longhi}(2022{\natexlab{a}})}]{Longhi2022}%
  \BibitemOpen
  \bibfield  {author} {\bibinfo {author} {\bibfnamefont {S.}~\bibnamefont
  {Longhi}},\ }\href {https://doi.org/10.1103/PhysRevLett.128.157601}
  {\bibfield  {journal} {\bibinfo  {journal} {Phys. Rev. Lett.}\ }\textbf
  {\bibinfo {volume} {128}},\ \bibinfo {pages} {157601} (\bibinfo {year}
  {2022}{\natexlab{a}})}\BibitemShut {NoStop}%
\bibitem [{\citenamefont {Longhi}(2022{\natexlab{b}})}]{Longhi2022v2}%
  \BibitemOpen
  \bibfield  {author} {\bibinfo {author} {\bibfnamefont {S.}~\bibnamefont
  {Longhi}},\ }\href {https://doi.org/10.1103/PhysRevB.105.245143} {\bibfield
  {journal} {\bibinfo  {journal} {Phys. Rev. B}\ }\textbf {\bibinfo {volume}
  {105}},\ \bibinfo {pages} {245143} (\bibinfo {year}
  {2022}{\natexlab{b}})}\BibitemShut {NoStop}%
\bibitem [{\citenamefont {Brandenbourger}\ \emph {et~al.}(2019)\citenamefont
  {Brandenbourger}, \citenamefont {Locsin}, \citenamefont {Lerner},\ and\
  \citenamefont {Coulais}}]{Brandenbourger2019}%
  \BibitemOpen
  \bibfield  {author} {\bibinfo {author} {\bibfnamefont {M.}~\bibnamefont
  {Brandenbourger}}, \bibinfo {author} {\bibfnamefont {X.}~\bibnamefont
  {Locsin}}, \bibinfo {author} {\bibfnamefont {E.}~\bibnamefont {Lerner}},\
  and\ \bibinfo {author} {\bibfnamefont {C.}~\bibnamefont {Coulais}},\
  }\href@noop {} {\bibfield  {journal} {\bibinfo  {journal} {Nat. Commun.}\
  }\textbf {\bibinfo {volume} {10}},\ \bibinfo {pages} {4608} (\bibinfo {year}
  {2019})}\BibitemShut {NoStop}%
\bibitem [{\citenamefont {Xiao}\ \emph {et~al.}(2020)\citenamefont {Xiao},
  \citenamefont {Deng}, \citenamefont {Wang}, \citenamefont {Zhu},
  \citenamefont {Wang}, \citenamefont {Yi},\ and\ \citenamefont
  {Xue}}]{Xiao2020}%
  \BibitemOpen
  \bibfield  {author} {\bibinfo {author} {\bibfnamefont {L.}~\bibnamefont
  {Xiao}}, \bibinfo {author} {\bibfnamefont {T.}~\bibnamefont {Deng}}, \bibinfo
  {author} {\bibfnamefont {K.}~\bibnamefont {Wang}}, \bibinfo {author}
  {\bibfnamefont {G.}~\bibnamefont {Zhu}}, \bibinfo {author} {\bibfnamefont
  {Z.}~\bibnamefont {Wang}}, \bibinfo {author} {\bibfnamefont {W.}~\bibnamefont
  {Yi}},\ and\ \bibinfo {author} {\bibfnamefont {P.}~\bibnamefont {Xue}},\
  }\href@noop {} {\bibfield  {journal} {\bibinfo  {journal} {Nat. Phys.}\
  }\textbf {\bibinfo {volume} {16}},\ \bibinfo {pages} {761} (\bibinfo {year}
  {2020})}\BibitemShut {NoStop}%
\bibitem [{\citenamefont {Weidemann}\ \emph {et~al.}(2020)\citenamefont
  {Weidemann}, \citenamefont {Kremer}, \citenamefont {Helbig}, \citenamefont
  {Hofmann}, \citenamefont {Stegmaier}, \citenamefont {Greiter}, \citenamefont
  {Thomale},\ and\ \citenamefont {Szameit}}]{Weidemann2020}%
  \BibitemOpen
  \bibfield  {author} {\bibinfo {author} {\bibfnamefont {S.}~\bibnamefont
  {Weidemann}}, \bibinfo {author} {\bibfnamefont {M.}~\bibnamefont {Kremer}},
  \bibinfo {author} {\bibfnamefont {T.}~\bibnamefont {Helbig}}, \bibinfo
  {author} {\bibfnamefont {T.}~\bibnamefont {Hofmann}}, \bibinfo {author}
  {\bibfnamefont {A.}~\bibnamefont {Stegmaier}}, \bibinfo {author}
  {\bibfnamefont {M.}~\bibnamefont {Greiter}}, \bibinfo {author} {\bibfnamefont
  {R.}~\bibnamefont {Thomale}},\ and\ \bibinfo {author} {\bibfnamefont
  {A.}~\bibnamefont {Szameit}},\ }\href@noop {} {\bibfield  {journal} {\bibinfo
   {journal} {Science}\ }\textbf {\bibinfo {volume} {368}},\ \bibinfo {pages}
  {311} (\bibinfo {year} {2020})}\BibitemShut {NoStop}%
\bibitem [{\citenamefont {Helbig}\ \emph {et~al.}(2020)\citenamefont {Helbig},
  \citenamefont {Hofmann}, \citenamefont {Imhof}, \citenamefont {Abdelghany},
  \citenamefont {Kiessling}, \citenamefont {Molenkamp}, \citenamefont {Lee},
  \citenamefont {Szameit}, \citenamefont {Greiter},\ and\ \citenamefont
  {Thomale}}]{Helbig2020}%
  \BibitemOpen
  \bibfield  {author} {\bibinfo {author} {\bibfnamefont {T.}~\bibnamefont
  {Helbig}}, \bibinfo {author} {\bibfnamefont {T.}~\bibnamefont {Hofmann}},
  \bibinfo {author} {\bibfnamefont {S.}~\bibnamefont {Imhof}}, \bibinfo
  {author} {\bibfnamefont {M.}~\bibnamefont {Abdelghany}}, \bibinfo {author}
  {\bibfnamefont {T.}~\bibnamefont {Kiessling}}, \bibinfo {author}
  {\bibfnamefont {L.}~\bibnamefont {Molenkamp}}, \bibinfo {author}
  {\bibfnamefont {C.}~\bibnamefont {Lee}}, \bibinfo {author} {\bibfnamefont
  {A.}~\bibnamefont {Szameit}}, \bibinfo {author} {\bibfnamefont
  {M.}~\bibnamefont {Greiter}},\ and\ \bibinfo {author} {\bibfnamefont
  {R.}~\bibnamefont {Thomale}},\ }\href@noop {} {\bibfield  {journal} {\bibinfo
   {journal} {Nat. Phys.}\ }\textbf {\bibinfo {volume} {16}},\ \bibinfo {pages}
  {747} (\bibinfo {year} {2020})}\BibitemShut {NoStop}%
\bibitem [{\citenamefont {Hofmann}\ \emph {et~al.}(2020)\citenamefont
  {Hofmann}, \citenamefont {Helbig}, \citenamefont {Schindler}, \citenamefont
  {Salgo}, \citenamefont {Brzezi\ifmmode~\acute{n}\else \'{n}\fi{}ska},
  \citenamefont {Greiter}, \citenamefont {Kiessling}, \citenamefont {Wolf},
  \citenamefont {Vollhardt}, \citenamefont {Kaba\ifmmode~\check{s}\else
  \v{s}\fi{}i}, \citenamefont {Lee}, \citenamefont {Bilu\ifmmode \check{s}\else
  \v{s}\fi{}i\ifmmode~\acute{c}\else \'{c}\fi{}}, \citenamefont {Thomale},\
  and\ \citenamefont {Neupert}}]{Hofmann2020}%
  \BibitemOpen
  \bibfield  {author} {\bibinfo {author} {\bibfnamefont {T.}~\bibnamefont
  {Hofmann}}, \bibinfo {author} {\bibfnamefont {T.}~\bibnamefont {Helbig}},
  \bibinfo {author} {\bibfnamefont {F.}~\bibnamefont {Schindler}}, \bibinfo
  {author} {\bibfnamefont {N.}~\bibnamefont {Salgo}}, \bibinfo {author}
  {\bibfnamefont {M.}~\bibnamefont {Brzezi\ifmmode~\acute{n}\else
  \'{n}\fi{}ska}}, \bibinfo {author} {\bibfnamefont {M.}~\bibnamefont
  {Greiter}}, \bibinfo {author} {\bibfnamefont {T.}~\bibnamefont {Kiessling}},
  \bibinfo {author} {\bibfnamefont {D.}~\bibnamefont {Wolf}}, \bibinfo {author}
  {\bibfnamefont {A.}~\bibnamefont {Vollhardt}}, \bibinfo {author}
  {\bibfnamefont {A.}~\bibnamefont {Kaba\ifmmode~\check{s}\else \v{s}\fi{}i}},
  \bibinfo {author} {\bibfnamefont {C.~H.}\ \bibnamefont {Lee}}, \bibinfo
  {author} {\bibfnamefont {A.}~\bibnamefont {Bilu\ifmmode \check{s}\else
  \v{s}\fi{}i\ifmmode~\acute{c}\else \'{c}\fi{}}}, \bibinfo {author}
  {\bibfnamefont {R.}~\bibnamefont {Thomale}},\ and\ \bibinfo {author}
  {\bibfnamefont {T.}~\bibnamefont {Neupert}},\ }\href
  {https://doi.org/10.1103/PhysRevResearch.2.023265} {\bibfield  {journal}
  {\bibinfo  {journal} {Phys. Rev. Research}\ }\textbf {\bibinfo {volume}
  {2}},\ \bibinfo {pages} {023265} (\bibinfo {year} {2020})}\BibitemShut
  {NoStop}%
\bibitem [{\citenamefont {Ghatak}\ \emph {et~al.}(2020)\citenamefont {Ghatak},
  \citenamefont {Brandenbourger}, \citenamefont {van Wezel},\ and\
  \citenamefont {Coulais}}]{Ghatak2020}%
  \BibitemOpen
  \bibfield  {author} {\bibinfo {author} {\bibfnamefont {A.}~\bibnamefont
  {Ghatak}}, \bibinfo {author} {\bibfnamefont {M.}~\bibnamefont
  {Brandenbourger}}, \bibinfo {author} {\bibfnamefont {J.}~\bibnamefont {van
  Wezel}},\ and\ \bibinfo {author} {\bibfnamefont {C.}~\bibnamefont
  {Coulais}},\ }\href@noop {} {\bibfield  {journal} {\bibinfo  {journal} {Proc.
  Nat. Ac. Sc. USA}\ }\textbf {\bibinfo {volume} {117}},\ \bibinfo {pages}
  {29561} (\bibinfo {year} {2020})}\BibitemShut {NoStop}%
\bibitem [{\citenamefont {Zhang}\ \emph
  {et~al.}(2021{\natexlab{a}})\citenamefont {Zhang}, \citenamefont {Tian},
  \citenamefont {Jiang}, \citenamefont {Lu},\ and\ \citenamefont
  {Chen}}]{XZhang2021}%
  \BibitemOpen
  \bibfield  {author} {\bibinfo {author} {\bibfnamefont {X.}~\bibnamefont
  {Zhang}}, \bibinfo {author} {\bibfnamefont {Y.}~\bibnamefont {Tian}},
  \bibinfo {author} {\bibfnamefont {J.-H.}\ \bibnamefont {Jiang}}, \bibinfo
  {author} {\bibfnamefont {M.-H.}\ \bibnamefont {Lu}},\ and\ \bibinfo {author}
  {\bibfnamefont {Y.-F.}\ \bibnamefont {Chen}},\ }\href@noop {} {\bibfield
  {journal} {\bibinfo  {journal} {Nat. Commun.}\ }\textbf {\bibinfo {volume}
  {12}},\ \bibinfo {pages} {5377} (\bibinfo {year}
  {2021}{\natexlab{a}})}\BibitemShut {NoStop}%
\bibitem [{\citenamefont {Chen}\ \emph {et~al.}(2021)\citenamefont {Chen},
  \citenamefont {Li}, \citenamefont {Scheibner}, \citenamefont {Vitelli},\ and\
  \citenamefont {Huang}}]{Chen2021}%
  \BibitemOpen
  \bibfield  {author} {\bibinfo {author} {\bibfnamefont {Y.}~\bibnamefont
  {Chen}}, \bibinfo {author} {\bibfnamefont {X.}~\bibnamefont {Li}}, \bibinfo
  {author} {\bibfnamefont {C.}~\bibnamefont {Scheibner}}, \bibinfo {author}
  {\bibfnamefont {V.}~\bibnamefont {Vitelli}},\ and\ \bibinfo {author}
  {\bibfnamefont {G.}~\bibnamefont {Huang}},\ }\href@noop {} {\bibfield
  {journal} {\bibinfo  {journal} {Nat. Commun.}\ }\textbf {\bibinfo {volume}
  {12}},\ \bibinfo {pages} {5935} (\bibinfo {year} {2021})}\BibitemShut
  {NoStop}%
\bibitem [{\citenamefont {Zhang}\ \emph
  {et~al.}(2021{\natexlab{b}})\citenamefont {Zhang}, \citenamefont {Yang},
  \citenamefont {Ge}, \citenamefont {Guan}, \citenamefont {Chen}, \citenamefont
  {Yan}, \citenamefont {Chen}, \citenamefont {Xi}, \citenamefont {Li},
  \citenamefont {Jia} \emph {et~al.}}]{Zhang2021}%
  \BibitemOpen
  \bibfield  {author} {\bibinfo {author} {\bibfnamefont {L.}~\bibnamefont
  {Zhang}}, \bibinfo {author} {\bibfnamefont {Y.}~\bibnamefont {Yang}},
  \bibinfo {author} {\bibfnamefont {Y.}~\bibnamefont {Ge}}, \bibinfo {author}
  {\bibfnamefont {Y.-J.}\ \bibnamefont {Guan}}, \bibinfo {author}
  {\bibfnamefont {Q.}~\bibnamefont {Chen}}, \bibinfo {author} {\bibfnamefont
  {Q.}~\bibnamefont {Yan}}, \bibinfo {author} {\bibfnamefont {F.}~\bibnamefont
  {Chen}}, \bibinfo {author} {\bibfnamefont {R.}~\bibnamefont {Xi}}, \bibinfo
  {author} {\bibfnamefont {Y.}~\bibnamefont {Li}}, \bibinfo {author}
  {\bibfnamefont {D.}~\bibnamefont {Jia}}, \emph {et~al.},\ }\href@noop {}
  {\bibfield  {journal} {\bibinfo  {journal} {Nat. Commun.}\ }\textbf {\bibinfo
  {volume} {12}},\ \bibinfo {pages} {6297} (\bibinfo {year}
  {2021}{\natexlab{b}})}\BibitemShut {NoStop}%
\bibitem [{\citenamefont {Wang}\ \emph {et~al.}(2022)\citenamefont {Wang},
  \citenamefont {Wang},\ and\ \citenamefont {Ma}}]{Wang2022}%
  \BibitemOpen
  \bibfield  {author} {\bibinfo {author} {\bibfnamefont {W.}~\bibnamefont
  {Wang}}, \bibinfo {author} {\bibfnamefont {X.}~\bibnamefont {Wang}},\ and\
  \bibinfo {author} {\bibfnamefont {G.}~\bibnamefont {Ma}},\ }\href@noop {}
  {\bibfield  {journal} {\bibinfo  {journal} {Nature}\ }\textbf {\bibinfo
  {volume} {608}},\ \bibinfo {pages} {50} (\bibinfo {year} {2022})}\BibitemShut
  {NoStop}%
\bibitem [{\citenamefont {Liang}\ \emph {et~al.}(2022)\citenamefont {Liang},
  \citenamefont {Xie}, \citenamefont {Dong}, \citenamefont {Li}, \citenamefont
  {Li}, \citenamefont {Gadway}, \citenamefont {Yi},\ and\ \citenamefont
  {Yan}}]{Liang2022}%
  \BibitemOpen
  \bibfield  {author} {\bibinfo {author} {\bibfnamefont {Q.}~\bibnamefont
  {Liang}}, \bibinfo {author} {\bibfnamefont {D.}~\bibnamefont {Xie}}, \bibinfo
  {author} {\bibfnamefont {Z.}~\bibnamefont {Dong}}, \bibinfo {author}
  {\bibfnamefont {H.}~\bibnamefont {Li}}, \bibinfo {author} {\bibfnamefont
  {H.}~\bibnamefont {Li}}, \bibinfo {author} {\bibfnamefont {B.}~\bibnamefont
  {Gadway}}, \bibinfo {author} {\bibfnamefont {W.}~\bibnamefont {Yi}},\ and\
  \bibinfo {author} {\bibfnamefont {B.}~\bibnamefont {Yan}},\ }\href
  {https://doi.org/10.1103/PhysRevLett.129.070401} {\bibfield  {journal}
  {\bibinfo  {journal} {Phys. Rev. Lett.}\ }\textbf {\bibinfo {volume} {129}},\
  \bibinfo {pages} {070401} (\bibinfo {year} {2022})}\BibitemShut {NoStop}%
\bibitem [{\citenamefont {Liu}\ \emph {et~al.}(2022)\citenamefont {Liu},
  \citenamefont {Wei}, \citenamefont {Hemmatyar}, \citenamefont {Pyrialakos},
  \citenamefont {Jung}, \citenamefont {Christodoulides},\ and\ \citenamefont
  {Khajavikhan}}]{Liu2022}%
  \BibitemOpen
  \bibfield  {author} {\bibinfo {author} {\bibfnamefont {Y.~G.}\ \bibnamefont
  {Liu}}, \bibinfo {author} {\bibfnamefont {Y.}~\bibnamefont {Wei}}, \bibinfo
  {author} {\bibfnamefont {O.}~\bibnamefont {Hemmatyar}}, \bibinfo {author}
  {\bibfnamefont {G.~G.}\ \bibnamefont {Pyrialakos}}, \bibinfo {author}
  {\bibfnamefont {P.~S.}\ \bibnamefont {Jung}}, \bibinfo {author}
  {\bibfnamefont {D.~N.}\ \bibnamefont {Christodoulides}},\ and\ \bibinfo
  {author} {\bibfnamefont {M.}~\bibnamefont {Khajavikhan}},\ }\href@noop {}
  {\bibfield  {journal} {\bibinfo  {journal} {Light Sci. Appl.}\ }\textbf
  {\bibinfo {volume} {11}},\ \bibinfo {pages} {336} (\bibinfo {year}
  {2022})}\BibitemShut {NoStop}%
\bibitem [{\citenamefont {Yokomizo}\ and\ \citenamefont
  {Murakami}(2019)}]{Yokomizo2019}%
  \BibitemOpen
  \bibfield  {author} {\bibinfo {author} {\bibfnamefont {K.}~\bibnamefont
  {Yokomizo}}\ and\ \bibinfo {author} {\bibfnamefont {S.}~\bibnamefont
  {Murakami}},\ }\href {https://doi.org/10.1103/PhysRevLett.123.066404}
  {\bibfield  {journal} {\bibinfo  {journal} {Phys. Rev. Lett.}\ }\textbf
  {\bibinfo {volume} {123}},\ \bibinfo {pages} {066404} (\bibinfo {year}
  {2019})}\BibitemShut {NoStop}%
\bibitem [{\citenamefont {Yokomizo}\ and\ \citenamefont
  {Murakami}(2020{\natexlab{a}})}]{Yokomizo2020}%
  \BibitemOpen
  \bibfield  {author} {\bibinfo {author} {\bibfnamefont {K.}~\bibnamefont
  {Yokomizo}}\ and\ \bibinfo {author} {\bibfnamefont {S.}~\bibnamefont
  {Murakami}},\ }\href@noop {} {\bibfield  {journal} {\bibinfo  {journal}
  {Prog. Theor. Exp. Phys.}\ }\textbf {\bibinfo {volume} {2020}},\ \bibinfo
  {pages} {12A102} (\bibinfo {year} {2020}{\natexlab{a}})}\BibitemShut
  {NoStop}%
\bibitem [{\citenamefont {Yokomizo}\ and\ \citenamefont
  {Murakami}(2021{\natexlab{b}})}]{Yokomizo2021}%
  \BibitemOpen
  \bibfield  {author} {\bibinfo {author} {\bibfnamefont {K.}~\bibnamefont
  {Yokomizo}}\ and\ \bibinfo {author} {\bibfnamefont {S.}~\bibnamefont
  {Murakami}},\ }\href {https://doi.org/10.1103/PhysRevB.103.165123} {\bibfield
   {journal} {\bibinfo  {journal} {Phys. Rev. B}\ }\textbf {\bibinfo {volume}
  {103}},\ \bibinfo {pages} {165123} (\bibinfo {year}
  {2021}{\natexlab{b}})}\BibitemShut {NoStop}%
\bibitem [{\citenamefont {Yokomizo}\ \emph {et~al.}(2022)\citenamefont
  {Yokomizo}, \citenamefont {Yoda},\ and\ \citenamefont
  {Murakami}}]{Yokomizo2022}%
  \BibitemOpen
  \bibfield  {author} {\bibinfo {author} {\bibfnamefont {K.}~\bibnamefont
  {Yokomizo}}, \bibinfo {author} {\bibfnamefont {T.}~\bibnamefont {Yoda}},\
  and\ \bibinfo {author} {\bibfnamefont {S.}~\bibnamefont {Murakami}},\ }\href
  {https://doi.org/10.1103/PhysRevResearch.4.023089} {\bibfield  {journal}
  {\bibinfo  {journal} {Phys. Rev. Research}\ }\textbf {\bibinfo {volume}
  {4}},\ \bibinfo {pages} {023089} (\bibinfo {year} {2022})}\BibitemShut
  {NoStop}%
\bibitem [{\citenamefont {Yokomizo}\ and\ \citenamefont
  {Murakami}(2020{\natexlab{b}})}]{Yokomizo2020v2}%
  \BibitemOpen
  \bibfield  {author} {\bibinfo {author} {\bibfnamefont {K.}~\bibnamefont
  {Yokomizo}}\ and\ \bibinfo {author} {\bibfnamefont {S.}~\bibnamefont
  {Murakami}},\ }\href {https://doi.org/10.1103/PhysRevResearch.2.043045}
  {\bibfield  {journal} {\bibinfo  {journal} {Phys. Rev. Research}\ }\textbf
  {\bibinfo {volume} {2}},\ \bibinfo {pages} {043045} (\bibinfo {year}
  {2020}{\natexlab{b}})}\BibitemShut {NoStop}%
\bibitem [{\citenamefont {Leykam}\ \emph {et~al.}(2017)\citenamefont {Leykam},
  \citenamefont {Bliokh}, \citenamefont {Huang}, \citenamefont {Chong},\ and\
  \citenamefont {Nori}}]{Leykam2017}%
  \BibitemOpen
  \bibfield  {author} {\bibinfo {author} {\bibfnamefont {D.}~\bibnamefont
  {Leykam}}, \bibinfo {author} {\bibfnamefont {K.~Y.}\ \bibnamefont {Bliokh}},
  \bibinfo {author} {\bibfnamefont {C.}~\bibnamefont {Huang}}, \bibinfo
  {author} {\bibfnamefont {Y.~D.}\ \bibnamefont {Chong}},\ and\ \bibinfo
  {author} {\bibfnamefont {F.}~\bibnamefont {Nori}},\ }\href
  {https://doi.org/10.1103/PhysRevLett.118.040401} {\bibfield  {journal}
  {\bibinfo  {journal} {Phys. Rev. Lett.}\ }\textbf {\bibinfo {volume} {118}},\
  \bibinfo {pages} {040401} (\bibinfo {year} {2017})}\BibitemShut {NoStop}%
\bibitem [{\citenamefont {Shen}\ \emph {et~al.}(2018)\citenamefont {Shen},
  \citenamefont {Zhen},\ and\ \citenamefont {Fu}}]{Shen2018}%
  \BibitemOpen
  \bibfield  {author} {\bibinfo {author} {\bibfnamefont {H.}~\bibnamefont
  {Shen}}, \bibinfo {author} {\bibfnamefont {B.}~\bibnamefont {Zhen}},\ and\
  \bibinfo {author} {\bibfnamefont {L.}~\bibnamefont {Fu}},\ }\href
  {https://doi.org/10.1103/PhysRevLett.120.146402} {\bibfield  {journal}
  {\bibinfo  {journal} {Phys. Rev. Lett.}\ }\textbf {\bibinfo {volume} {120}},\
  \bibinfo {pages} {146402} (\bibinfo {year} {2018})}\BibitemShut {NoStop}%
\bibitem [{\citenamefont {Kawabata}\ \emph {et~al.}(2018)\citenamefont
  {Kawabata}, \citenamefont {Shiozaki},\ and\ \citenamefont
  {Ueda}}]{Kawabata2018}%
  \BibitemOpen
  \bibfield  {author} {\bibinfo {author} {\bibfnamefont {K.}~\bibnamefont
  {Kawabata}}, \bibinfo {author} {\bibfnamefont {K.}~\bibnamefont {Shiozaki}},\
  and\ \bibinfo {author} {\bibfnamefont {M.}~\bibnamefont {Ueda}},\ }\href
  {https://doi.org/10.1103/PhysRevB.98.165148} {\bibfield  {journal} {\bibinfo
  {journal} {Phys. Rev. B}\ }\textbf {\bibinfo {volume} {98}},\ \bibinfo
  {pages} {165148} (\bibinfo {year} {2018})}\BibitemShut {NoStop}%
\bibitem [{\citenamefont {Ezawa}(2019)}]{Ezawa2019}%
  \BibitemOpen
  \bibfield  {author} {\bibinfo {author} {\bibfnamefont {M.}~\bibnamefont
  {Ezawa}},\ }\href {https://doi.org/10.1103/PhysRevB.100.081401} {\bibfield
  {journal} {\bibinfo  {journal} {Phys. Rev. B}\ }\textbf {\bibinfo {volume}
  {100}},\ \bibinfo {pages} {081401} (\bibinfo {year} {2019})}\BibitemShut
  {NoStop}%
\bibitem [{\citenamefont {Luo}\ and\ \citenamefont {Zhang}(2019)}]{Luo2019}%
  \BibitemOpen
  \bibfield  {author} {\bibinfo {author} {\bibfnamefont {X.-W.}\ \bibnamefont
  {Luo}}\ and\ \bibinfo {author} {\bibfnamefont {C.}~\bibnamefont {Zhang}},\
  }\href {https://doi.org/10.1103/PhysRevLett.123.073601} {\bibfield  {journal}
  {\bibinfo  {journal} {Phys. Rev. Lett.}\ }\textbf {\bibinfo {volume} {123}},\
  \bibinfo {pages} {073601} (\bibinfo {year} {2019})}\BibitemShut {NoStop}%
\bibitem [{\citenamefont {Wu}\ \emph {et~al.}(2019)\citenamefont {Wu},
  \citenamefont {Jin},\ and\ \citenamefont {Song}}]{Wu2019}%
  \BibitemOpen
  \bibfield  {author} {\bibinfo {author} {\bibfnamefont {H.~C.}\ \bibnamefont
  {Wu}}, \bibinfo {author} {\bibfnamefont {L.}~\bibnamefont {Jin}},\ and\
  \bibinfo {author} {\bibfnamefont {Z.}~\bibnamefont {Song}},\ }\href
  {https://doi.org/10.1103/PhysRevB.100.155117} {\bibfield  {journal} {\bibinfo
   {journal} {Phys. Rev. B}\ }\textbf {\bibinfo {volume} {100}},\ \bibinfo
  {pages} {155117} (\bibinfo {year} {2019})}\BibitemShut {NoStop}%
\bibitem [{\citenamefont {Wu}\ \emph {et~al.}(2020)\citenamefont {Wu},
  \citenamefont {Liu},\ and\ \citenamefont {Hou}}]{Wu2020}%
  \BibitemOpen
  \bibfield  {author} {\bibinfo {author} {\bibfnamefont {Y.-J.}\ \bibnamefont
  {Wu}}, \bibinfo {author} {\bibfnamefont {C.-C.}\ \bibnamefont {Liu}},\ and\
  \bibinfo {author} {\bibfnamefont {J.}~\bibnamefont {Hou}},\ }\href
  {https://doi.org/10.1103/PhysRevA.101.043833} {\bibfield  {journal} {\bibinfo
   {journal} {Phys. Rev. A}\ }\textbf {\bibinfo {volume} {101}},\ \bibinfo
  {pages} {043833} (\bibinfo {year} {2020})}\BibitemShut {NoStop}%
\bibitem [{\citenamefont {Xue}\ \emph {et~al.}(2020)\citenamefont {Xue},
  \citenamefont {Wang}, \citenamefont {Zhang},\ and\ \citenamefont
  {Chong}}]{Xue2020}%
  \BibitemOpen
  \bibfield  {author} {\bibinfo {author} {\bibfnamefont {H.}~\bibnamefont
  {Xue}}, \bibinfo {author} {\bibfnamefont {Q.}~\bibnamefont {Wang}}, \bibinfo
  {author} {\bibfnamefont {B.}~\bibnamefont {Zhang}},\ and\ \bibinfo {author}
  {\bibfnamefont {Y.~D.}\ \bibnamefont {Chong}},\ }\href
  {https://doi.org/10.1103/PhysRevLett.124.236403} {\bibfield  {journal}
  {\bibinfo  {journal} {Phys. Rev. Lett.}\ }\textbf {\bibinfo {volume} {124}},\
  \bibinfo {pages} {236403} (\bibinfo {year} {2020})}\BibitemShut {NoStop}%
\bibitem [{\citenamefont {Ao}\ \emph {et~al.}(2020)\citenamefont {Ao},
  \citenamefont {Hu}, \citenamefont {You}, \citenamefont {Lu}, \citenamefont
  {Fu}, \citenamefont {Wang},\ and\ \citenamefont {Gong}}]{Ao2020}%
  \BibitemOpen
  \bibfield  {author} {\bibinfo {author} {\bibfnamefont {Y.}~\bibnamefont
  {Ao}}, \bibinfo {author} {\bibfnamefont {X.}~\bibnamefont {Hu}}, \bibinfo
  {author} {\bibfnamefont {Y.}~\bibnamefont {You}}, \bibinfo {author}
  {\bibfnamefont {C.}~\bibnamefont {Lu}}, \bibinfo {author} {\bibfnamefont
  {Y.}~\bibnamefont {Fu}}, \bibinfo {author} {\bibfnamefont {X.}~\bibnamefont
  {Wang}},\ and\ \bibinfo {author} {\bibfnamefont {Q.}~\bibnamefont {Gong}},\
  }\href {https://doi.org/10.1103/PhysRevLett.125.013902} {\bibfield  {journal}
  {\bibinfo  {journal} {Phys. Rev. Lett.}\ }\textbf {\bibinfo {volume} {125}},\
  \bibinfo {pages} {013902} (\bibinfo {year} {2020})}\BibitemShut {NoStop}%
\bibitem [{\citenamefont {Song}\ \emph
  {et~al.}(2020{\natexlab{a}})\citenamefont {Song}, \citenamefont {Sun},
  \citenamefont {Dutt}, \citenamefont {Minkov}, \citenamefont {Wojcik},
  \citenamefont {Wang}, \citenamefont {Williamson}, \citenamefont {Orenstein},\
  and\ \citenamefont {Fan}}]{Song2020}%
  \BibitemOpen
  \bibfield  {author} {\bibinfo {author} {\bibfnamefont {A.~Y.}\ \bibnamefont
  {Song}}, \bibinfo {author} {\bibfnamefont {X.-Q.}\ \bibnamefont {Sun}},
  \bibinfo {author} {\bibfnamefont {A.}~\bibnamefont {Dutt}}, \bibinfo {author}
  {\bibfnamefont {M.}~\bibnamefont {Minkov}}, \bibinfo {author} {\bibfnamefont
  {C.}~\bibnamefont {Wojcik}}, \bibinfo {author} {\bibfnamefont
  {H.}~\bibnamefont {Wang}}, \bibinfo {author} {\bibfnamefont {I.~A.~D.}\
  \bibnamefont {Williamson}}, \bibinfo {author} {\bibfnamefont
  {M.}~\bibnamefont {Orenstein}},\ and\ \bibinfo {author} {\bibfnamefont
  {S.}~\bibnamefont {Fan}},\ }\href
  {https://doi.org/10.1103/PhysRevLett.125.033603} {\bibfield  {journal}
  {\bibinfo  {journal} {Phys. Rev. Lett.}\ }\textbf {\bibinfo {volume} {125}},\
  \bibinfo {pages} {033603} (\bibinfo {year} {2020}{\natexlab{a}})}\BibitemShut
  {NoStop}%
\bibitem [{\citenamefont {Xie}\ \emph {et~al.}(2021)\citenamefont {Xie},
  \citenamefont {Wu}, \citenamefont {Zhang}, \citenamefont {Jin},\ and\
  \citenamefont {Song}}]{Xie2021}%
  \BibitemOpen
  \bibfield  {author} {\bibinfo {author} {\bibfnamefont {L.~C.}\ \bibnamefont
  {Xie}}, \bibinfo {author} {\bibfnamefont {H.~C.}\ \bibnamefont {Wu}},
  \bibinfo {author} {\bibfnamefont {X.~Z.}\ \bibnamefont {Zhang}}, \bibinfo
  {author} {\bibfnamefont {L.}~\bibnamefont {Jin}},\ and\ \bibinfo {author}
  {\bibfnamefont {Z.}~\bibnamefont {Song}},\ }\href
  {https://doi.org/10.1103/PhysRevB.104.125406} {\bibfield  {journal} {\bibinfo
   {journal} {Phys. Rev. B}\ }\textbf {\bibinfo {volume} {104}},\ \bibinfo
  {pages} {125406} (\bibinfo {year} {2021})}\BibitemShut {NoStop}%
\bibitem [{\citenamefont {Teo}\ \emph {et~al.}(2022)\citenamefont {Teo},
  \citenamefont {Xue},\ and\ \citenamefont {Zhang}}]{Teo2022}%
  \BibitemOpen
  \bibfield  {author} {\bibinfo {author} {\bibfnamefont {H.~T.}\ \bibnamefont
  {Teo}}, \bibinfo {author} {\bibfnamefont {H.}~\bibnamefont {Xue}},\ and\
  \bibinfo {author} {\bibfnamefont {B.}~\bibnamefont {Zhang}},\ }\href
  {https://doi.org/10.1103/PhysRevA.105.053510} {\bibfield  {journal} {\bibinfo
   {journal} {Phys. Rev. A}\ }\textbf {\bibinfo {volume} {105}},\ \bibinfo
  {pages} {053510} (\bibinfo {year} {2022})}\BibitemShut {NoStop}%
\bibitem [{\citenamefont {Philip}\ \emph {et~al.}(2018)\citenamefont {Philip},
  \citenamefont {Hirsbrunner},\ and\ \citenamefont {Gilbert}}]{Philip2018}%
  \BibitemOpen
  \bibfield  {author} {\bibinfo {author} {\bibfnamefont {T.~M.}\ \bibnamefont
  {Philip}}, \bibinfo {author} {\bibfnamefont {M.~R.}\ \bibnamefont
  {Hirsbrunner}},\ and\ \bibinfo {author} {\bibfnamefont {M.~J.}\ \bibnamefont
  {Gilbert}},\ }\href {https://doi.org/10.1103/PhysRevB.98.155430} {\bibfield
  {journal} {\bibinfo  {journal} {Phys. Rev. B}\ }\textbf {\bibinfo {volume}
  {98}},\ \bibinfo {pages} {155430} (\bibinfo {year} {2018})}\BibitemShut
  {NoStop}%
\bibitem [{\citenamefont {Chen}\ and\ \citenamefont {Zhai}(2018)}]{Chen2018}%
  \BibitemOpen
  \bibfield  {author} {\bibinfo {author} {\bibfnamefont {Y.}~\bibnamefont
  {Chen}}\ and\ \bibinfo {author} {\bibfnamefont {H.}~\bibnamefont {Zhai}},\
  }\href {https://doi.org/10.1103/PhysRevB.98.245130} {\bibfield  {journal}
  {\bibinfo  {journal} {Phys. Rev. B}\ }\textbf {\bibinfo {volume} {98}},\
  \bibinfo {pages} {245130} (\bibinfo {year} {2018})}\BibitemShut {NoStop}%
\bibitem [{\citenamefont {Hirsbrunner}\ \emph {et~al.}(2019)\citenamefont
  {Hirsbrunner}, \citenamefont {Philip},\ and\ \citenamefont
  {Gilbert}}]{Hirsbrunner2019}%
  \BibitemOpen
  \bibfield  {author} {\bibinfo {author} {\bibfnamefont {M.~R.}\ \bibnamefont
  {Hirsbrunner}}, \bibinfo {author} {\bibfnamefont {T.~M.}\ \bibnamefont
  {Philip}},\ and\ \bibinfo {author} {\bibfnamefont {M.~J.}\ \bibnamefont
  {Gilbert}},\ }\href {https://doi.org/10.1103/PhysRevB.100.081104} {\bibfield
  {journal} {\bibinfo  {journal} {Phys. Rev. B}\ }\textbf {\bibinfo {volume}
  {100}},\ \bibinfo {pages} {081104} (\bibinfo {year} {2019})}\BibitemShut
  {NoStop}%
\bibitem [{\citenamefont {Groenendijk}\ \emph {et~al.}(2021)\citenamefont
  {Groenendijk}, \citenamefont {Schmidt},\ and\ \citenamefont
  {Meng}}]{Groenendijk2021}%
  \BibitemOpen
  \bibfield  {author} {\bibinfo {author} {\bibfnamefont {S.}~\bibnamefont
  {Groenendijk}}, \bibinfo {author} {\bibfnamefont {T.~L.}\ \bibnamefont
  {Schmidt}},\ and\ \bibinfo {author} {\bibfnamefont {T.}~\bibnamefont
  {Meng}},\ }\href {https://doi.org/10.1103/PhysRevResearch.3.023001}
  {\bibfield  {journal} {\bibinfo  {journal} {Phys. Rev. Research}\ }\textbf
  {\bibinfo {volume} {3}},\ \bibinfo {pages} {023001} (\bibinfo {year}
  {2021})}\BibitemShut {NoStop}%
\bibitem [{\citenamefont {Lee}\ \emph {et~al.}(2019)\citenamefont {Lee},
  \citenamefont {Li},\ and\ \citenamefont {Gong}}]{Lee2019}%
  \BibitemOpen
  \bibfield  {author} {\bibinfo {author} {\bibfnamefont {C.~H.}\ \bibnamefont
  {Lee}}, \bibinfo {author} {\bibfnamefont {L.}~\bibnamefont {Li}},\ and\
  \bibinfo {author} {\bibfnamefont {J.}~\bibnamefont {Gong}},\ }\href
  {https://doi.org/10.1103/PhysRevLett.123.016805} {\bibfield  {journal}
  {\bibinfo  {journal} {Phys. Rev. Lett.}\ }\textbf {\bibinfo {volume} {123}},\
  \bibinfo {pages} {016805} (\bibinfo {year} {2019})}\BibitemShut {NoStop}%
\bibitem [{\citenamefont {Kawabata}\ \emph {et~al.}(2020)\citenamefont
  {Kawabata}, \citenamefont {Sato},\ and\ \citenamefont
  {Shiozaki}}]{Kawabata2020}%
  \BibitemOpen
  \bibfield  {author} {\bibinfo {author} {\bibfnamefont {K.}~\bibnamefont
  {Kawabata}}, \bibinfo {author} {\bibfnamefont {M.}~\bibnamefont {Sato}},\
  and\ \bibinfo {author} {\bibfnamefont {K.}~\bibnamefont {Shiozaki}},\ }\href
  {https://doi.org/10.1103/PhysRevB.102.205118} {\bibfield  {journal} {\bibinfo
   {journal} {Phys. Rev. B}\ }\textbf {\bibinfo {volume} {102}},\ \bibinfo
  {pages} {205118} (\bibinfo {year} {2020})}\BibitemShut {NoStop}%
\bibitem [{\citenamefont {Okugawa}\ \emph {et~al.}(2020)\citenamefont
  {Okugawa}, \citenamefont {Takahashi},\ and\ \citenamefont
  {Yokomizo}}]{Okugawa2020}%
  \BibitemOpen
  \bibfield  {author} {\bibinfo {author} {\bibfnamefont {R.}~\bibnamefont
  {Okugawa}}, \bibinfo {author} {\bibfnamefont {R.}~\bibnamefont {Takahashi}},\
  and\ \bibinfo {author} {\bibfnamefont {K.}~\bibnamefont {Yokomizo}},\ }\href
  {https://doi.org/10.1103/PhysRevB.102.241202} {\bibfield  {journal} {\bibinfo
   {journal} {Phys. Rev. B}\ }\textbf {\bibinfo {volume} {102}},\ \bibinfo
  {pages} {241202} (\bibinfo {year} {2020})}\BibitemShut {NoStop}%
\bibitem [{\citenamefont {Fu}\ \emph {et~al.}(2021)\citenamefont {Fu},
  \citenamefont {Hu},\ and\ \citenamefont {Wan}}]{Fu2021}%
  \BibitemOpen
  \bibfield  {author} {\bibinfo {author} {\bibfnamefont {Y.}~\bibnamefont
  {Fu}}, \bibinfo {author} {\bibfnamefont {J.}~\bibnamefont {Hu}},\ and\
  \bibinfo {author} {\bibfnamefont {S.}~\bibnamefont {Wan}},\ }\href
  {https://doi.org/10.1103/PhysRevB.103.045420} {\bibfield  {journal} {\bibinfo
   {journal} {Phys. Rev. B}\ }\textbf {\bibinfo {volume} {103}},\ \bibinfo
  {pages} {045420} (\bibinfo {year} {2021})}\BibitemShut {NoStop}%
\bibitem [{\citenamefont {Zhang}\ \emph {et~al.}(2022)\citenamefont {Zhang},
  \citenamefont {Yang},\ and\ \citenamefont {Fang}}]{Zhang2022}%
  \BibitemOpen
  \bibfield  {author} {\bibinfo {author} {\bibfnamefont {K.}~\bibnamefont
  {Zhang}}, \bibinfo {author} {\bibfnamefont {Z.}~\bibnamefont {Yang}},\ and\
  \bibinfo {author} {\bibfnamefont {C.}~\bibnamefont {Fang}},\ }\href@noop {}
  {\bibfield  {journal} {\bibinfo  {journal} {Nat. Commun.}\ }\textbf {\bibinfo
  {volume} {13}},\ \bibinfo {pages} {2496} (\bibinfo {year}
  {2022})}\BibitemShut {NoStop}%
\bibitem [{\citenamefont {Li}\ \emph {et~al.}(2022)\citenamefont {Li},
  \citenamefont {Liang}, \citenamefont {Wang}, \citenamefont {Lu},\ and\
  \citenamefont {Liu}}]{Li2022}%
  \BibitemOpen
  \bibfield  {author} {\bibinfo {author} {\bibfnamefont {Y.}~\bibnamefont
  {Li}}, \bibinfo {author} {\bibfnamefont {C.}~\bibnamefont {Liang}}, \bibinfo
  {author} {\bibfnamefont {C.}~\bibnamefont {Wang}}, \bibinfo {author}
  {\bibfnamefont {C.}~\bibnamefont {Lu}},\ and\ \bibinfo {author}
  {\bibfnamefont {Y.-C.}\ \bibnamefont {Liu}},\ }\href
  {https://doi.org/10.1103/PhysRevLett.128.223903} {\bibfield  {journal}
  {\bibinfo  {journal} {Phys. Rev. Lett.}\ }\textbf {\bibinfo {volume} {128}},\
  \bibinfo {pages} {223903} (\bibinfo {year} {2022})}\BibitemShut {NoStop}%
\bibitem [{\citenamefont {Zhu}\ and\ \citenamefont {Gong}(2022)}]{Zhu2022}%
  \BibitemOpen
  \bibfield  {author} {\bibinfo {author} {\bibfnamefont {W.}~\bibnamefont
  {Zhu}}\ and\ \bibinfo {author} {\bibfnamefont {J.}~\bibnamefont {Gong}},\
  }\href {https://doi.org/10.1103/PhysRevB.106.035425} {\bibfield  {journal}
  {\bibinfo  {journal} {Phys. Rev. B}\ }\textbf {\bibinfo {volume} {106}},\
  \bibinfo {pages} {035425} (\bibinfo {year} {2022})}\BibitemShut {NoStop}%
\bibitem [{\citenamefont {Fang}\ \emph {et~al.}(2022)\citenamefont {Fang},
  \citenamefont {Hu}, \citenamefont {Zhou},\ and\ \citenamefont
  {Ding}}]{Fang2022}%
  \BibitemOpen
  \bibfield  {author} {\bibinfo {author} {\bibfnamefont {Z.}~\bibnamefont
  {Fang}}, \bibinfo {author} {\bibfnamefont {M.}~\bibnamefont {Hu}}, \bibinfo
  {author} {\bibfnamefont {L.}~\bibnamefont {Zhou}},\ and\ \bibinfo {author}
  {\bibfnamefont {K.}~\bibnamefont {Ding}},\ }\href@noop {} {\bibfield
  {journal} {\bibinfo  {journal} {Nanophotonics}\ }\textbf {\bibinfo {volume}
  {11}},\ \bibinfo {pages} {3447} (\bibinfo {year} {2022})}\BibitemShut
  {NoStop}%
\bibitem [{\citenamefont {Palacios}\ \emph {et~al.}(2021)\citenamefont
  {Palacios}, \citenamefont {Tchoumakov}, \citenamefont {Guix}, \citenamefont
  {Pagonabarraga}, \citenamefont {S{\'a}nchez},\ and\ \citenamefont
  {G~Grushin}}]{Palacios2021}%
  \BibitemOpen
  \bibfield  {author} {\bibinfo {author} {\bibfnamefont {L.~S.}\ \bibnamefont
  {Palacios}}, \bibinfo {author} {\bibfnamefont {S.}~\bibnamefont
  {Tchoumakov}}, \bibinfo {author} {\bibfnamefont {M.}~\bibnamefont {Guix}},
  \bibinfo {author} {\bibfnamefont {I.}~\bibnamefont {Pagonabarraga}}, \bibinfo
  {author} {\bibfnamefont {S.}~\bibnamefont {S{\'a}nchez}},\ and\ \bibinfo
  {author} {\bibfnamefont {A.}~\bibnamefont {G~Grushin}},\ }\href@noop {}
  {\bibfield  {journal} {\bibinfo  {journal} {Nat. Commun.}\ }\textbf {\bibinfo
  {volume} {12}},\ \bibinfo {pages} {4691} (\bibinfo {year}
  {2021})}\BibitemShut {NoStop}%
\bibitem [{\citenamefont {Zou}\ \emph {et~al.}(2021)\citenamefont {Zou},
  \citenamefont {Chen}, \citenamefont {He}, \citenamefont {Bao}, \citenamefont
  {Lee}, \citenamefont {Sun},\ and\ \citenamefont {Zhang}}]{Zou2021}%
  \BibitemOpen
  \bibfield  {author} {\bibinfo {author} {\bibfnamefont {D.}~\bibnamefont
  {Zou}}, \bibinfo {author} {\bibfnamefont {T.}~\bibnamefont {Chen}}, \bibinfo
  {author} {\bibfnamefont {W.}~\bibnamefont {He}}, \bibinfo {author}
  {\bibfnamefont {J.}~\bibnamefont {Bao}}, \bibinfo {author} {\bibfnamefont
  {C.~H.}\ \bibnamefont {Lee}}, \bibinfo {author} {\bibfnamefont
  {H.}~\bibnamefont {Sun}},\ and\ \bibinfo {author} {\bibfnamefont
  {X.}~\bibnamefont {Zhang}},\ }\href@noop {} {\bibfield  {journal} {\bibinfo
  {journal} {Nat. Commun.}\ }\textbf {\bibinfo {volume} {12}},\ \bibinfo
  {pages} {7201} (\bibinfo {year} {2021})}\BibitemShut {NoStop}%
\bibitem [{\citenamefont {Xu}\ \emph {et~al.}(2022)\citenamefont {Xu},
  \citenamefont {Bao},\ and\ \citenamefont {Liew}}]{Xu2022}%
  \BibitemOpen
  \bibfield  {author} {\bibinfo {author} {\bibfnamefont {X.}~\bibnamefont
  {Xu}}, \bibinfo {author} {\bibfnamefont {R.}~\bibnamefont {Bao}},\ and\
  \bibinfo {author} {\bibfnamefont {T.~C.~H.}\ \bibnamefont {Liew}},\ }\href
  {https://doi.org/10.1103/PhysRevB.106.L201302} {\bibfield  {journal}
  {\bibinfo  {journal} {Phys. Rev. B}\ }\textbf {\bibinfo {volume} {106}},\
  \bibinfo {pages} {L201302} (\bibinfo {year} {2022})}\BibitemShut {NoStop}%
\bibitem [{\citenamefont {Shang}\ \emph {et~al.}(2022)\citenamefont {Shang},
  \citenamefont {Liu}, \citenamefont {Shao}, \citenamefont {Han}, \citenamefont
  {Zang}, \citenamefont {Zhang}, \citenamefont {Salama}, \citenamefont {Gao},
  \citenamefont {Lee}, \citenamefont {Thomale}, \citenamefont {Manchon},
  \citenamefont {Zhang}, \citenamefont {Cui},\ and\ \citenamefont
  {Schwingenschl\"{o}gl}}]{Shang2022}%
  \BibitemOpen
  \bibfield  {author} {\bibinfo {author} {\bibfnamefont {C.}~\bibnamefont
  {Shang}}, \bibinfo {author} {\bibfnamefont {S.}~\bibnamefont {Liu}}, \bibinfo
  {author} {\bibfnamefont {R.}~\bibnamefont {Shao}}, \bibinfo {author}
  {\bibfnamefont {P.}~\bibnamefont {Han}}, \bibinfo {author} {\bibfnamefont
  {X.}~\bibnamefont {Zang}}, \bibinfo {author} {\bibfnamefont {X.}~\bibnamefont
  {Zhang}}, \bibinfo {author} {\bibfnamefont {K.~N.}\ \bibnamefont {Salama}},
  \bibinfo {author} {\bibfnamefont {W.}~\bibnamefont {Gao}}, \bibinfo {author}
  {\bibfnamefont {C.~H.}\ \bibnamefont {Lee}}, \bibinfo {author} {\bibfnamefont
  {R.}~\bibnamefont {Thomale}}, \bibinfo {author} {\bibfnamefont
  {A.}~\bibnamefont {Manchon}}, \bibinfo {author} {\bibfnamefont
  {S.}~\bibnamefont {Zhang}}, \bibinfo {author} {\bibfnamefont {T.~J.}\
  \bibnamefont {Cui}},\ and\ \bibinfo {author} {\bibfnamefont {U.}~\bibnamefont
  {Schwingenschl\"{o}gl}},\ }\href@noop {} {\bibfield  {journal} {\bibinfo
  {journal} {Advanced Science}\ }\textbf {\bibinfo {volume} {9}},\ \bibinfo
  {pages} {2202922} (\bibinfo {year} {2022})}\BibitemShut {NoStop}%
\bibitem [{\citenamefont {Wu}\ \emph {et~al.}(2023)\citenamefont {Wu},
  \citenamefont {Zhao}, \citenamefont {Kang}, \citenamefont {Weng},
  \citenamefont {Chi}, \citenamefont {Peng}, \citenamefont {Liu}, \citenamefont
  {Werner}, \citenamefont {Meng},\ and\ \citenamefont {Zhou}}]{Wu2022}%
  \BibitemOpen
  \bibfield  {author} {\bibinfo {author} {\bibfnamefont {M.}~\bibnamefont
  {Wu}}, \bibinfo {author} {\bibfnamefont {Q.}~\bibnamefont {Zhao}}, \bibinfo
  {author} {\bibfnamefont {L.}~\bibnamefont {Kang}}, \bibinfo {author}
  {\bibfnamefont {M.}~\bibnamefont {Weng}}, \bibinfo {author} {\bibfnamefont
  {Z.}~\bibnamefont {Chi}}, \bibinfo {author} {\bibfnamefont {R.}~\bibnamefont
  {Peng}}, \bibinfo {author} {\bibfnamefont {J.}~\bibnamefont {Liu}}, \bibinfo
  {author} {\bibfnamefont {D.~H.}\ \bibnamefont {Werner}}, \bibinfo {author}
  {\bibfnamefont {Y.}~\bibnamefont {Meng}},\ and\ \bibinfo {author}
  {\bibfnamefont {J.}~\bibnamefont {Zhou}},\ }\href
  {https://doi.org/10.1103/PhysRevB.107.064307} {\bibfield  {journal} {\bibinfo
   {journal} {Phys. Rev. B}\ }\textbf {\bibinfo {volume} {107}},\ \bibinfo
  {pages} {064307} (\bibinfo {year} {2023})}\BibitemShut {NoStop}%
\bibitem [{\citenamefont {Yao}\ \emph {et~al.}(2018)\citenamefont {Yao},
  \citenamefont {Song},\ and\ \citenamefont {Wang}}]{Yao2018v2}%
  \BibitemOpen
  \bibfield  {author} {\bibinfo {author} {\bibfnamefont {S.}~\bibnamefont
  {Yao}}, \bibinfo {author} {\bibfnamefont {F.}~\bibnamefont {Song}},\ and\
  \bibinfo {author} {\bibfnamefont {Z.}~\bibnamefont {Wang}},\ }\href
  {https://doi.org/10.1103/PhysRevLett.121.136802} {\bibfield  {journal}
  {\bibinfo  {journal} {Phys. Rev. Lett.}\ }\textbf {\bibinfo {volume} {121}},\
  \bibinfo {pages} {136802} (\bibinfo {year} {2018})}\BibitemShut {NoStop}%
\bibitem [{\citenamefont {Liu}\ \emph {et~al.}(2019)\citenamefont {Liu},
  \citenamefont {Zhang}, \citenamefont {Ai}, \citenamefont {Gong},
  \citenamefont {Kawabata}, \citenamefont {Ueda},\ and\ \citenamefont
  {Nori}}]{Liu2019}%
  \BibitemOpen
  \bibfield  {author} {\bibinfo {author} {\bibfnamefont {T.}~\bibnamefont
  {Liu}}, \bibinfo {author} {\bibfnamefont {Y.-R.}\ \bibnamefont {Zhang}},
  \bibinfo {author} {\bibfnamefont {Q.}~\bibnamefont {Ai}}, \bibinfo {author}
  {\bibfnamefont {Z.}~\bibnamefont {Gong}}, \bibinfo {author} {\bibfnamefont
  {K.}~\bibnamefont {Kawabata}}, \bibinfo {author} {\bibfnamefont
  {M.}~\bibnamefont {Ueda}},\ and\ \bibinfo {author} {\bibfnamefont
  {F.}~\bibnamefont {Nori}},\ }\href
  {https://doi.org/10.1103/PhysRevLett.122.076801} {\bibfield  {journal}
  {\bibinfo  {journal} {Phys. Rev. Lett.}\ }\textbf {\bibinfo {volume} {122}},\
  \bibinfo {pages} {076801} (\bibinfo {year} {2019})}\BibitemShut {NoStop}%
\bibitem [{\citenamefont {Yu}\ \emph {et~al.}(2021)\citenamefont {Yu},
  \citenamefont {Jung},\ and\ \citenamefont {Shvets}}]{Yu2021}%
  \BibitemOpen
  \bibfield  {author} {\bibinfo {author} {\bibfnamefont {Y.}~\bibnamefont
  {Yu}}, \bibinfo {author} {\bibfnamefont {M.}~\bibnamefont {Jung}},\ and\
  \bibinfo {author} {\bibfnamefont {G.}~\bibnamefont {Shvets}},\ }\href
  {https://doi.org/10.1103/PhysRevB.103.L041102} {\bibfield  {journal}
  {\bibinfo  {journal} {Phys. Rev. B}\ }\textbf {\bibinfo {volume} {103}},\
  \bibinfo {pages} {L041102} (\bibinfo {year} {2021})}\BibitemShut {NoStop}%
\bibitem [{\citenamefont {Lin}\ \emph {et~al.}(2021)\citenamefont {Lin},
  \citenamefont {Ding}, \citenamefont {Chen}, \citenamefont {Li}, \citenamefont
  {Ke}, \citenamefont {Li},\ and\ \citenamefont {Wang}}]{Lin2021}%
  \BibitemOpen
  \bibfield  {author} {\bibinfo {author} {\bibfnamefont {Z.}~\bibnamefont
  {Lin}}, \bibinfo {author} {\bibfnamefont {L.}~\bibnamefont {Ding}}, \bibinfo
  {author} {\bibfnamefont {S.}~\bibnamefont {Chen}}, \bibinfo {author}
  {\bibfnamefont {S.}~\bibnamefont {Li}}, \bibinfo {author} {\bibfnamefont
  {S.}~\bibnamefont {Ke}}, \bibinfo {author} {\bibfnamefont {X.}~\bibnamefont
  {Li}},\ and\ \bibinfo {author} {\bibfnamefont {B.}~\bibnamefont {Wang}},\
  }\href {https://doi.org/10.1103/PhysRevA.103.063507} {\bibfield  {journal}
  {\bibinfo  {journal} {Phys. Rev. A}\ }\textbf {\bibinfo {volume} {103}},\
  \bibinfo {pages} {063507} (\bibinfo {year} {2021})}\BibitemShut {NoStop}%
\bibitem [{\citenamefont {Xiao}\ and\ \citenamefont {Chan}(2022)}]{Xiao2022}%
  \BibitemOpen
  \bibfield  {author} {\bibinfo {author} {\bibfnamefont {Y.-X.}\ \bibnamefont
  {Xiao}}\ and\ \bibinfo {author} {\bibfnamefont {C.~T.}\ \bibnamefont
  {Chan}},\ }\href {https://doi.org/10.1103/PhysRevB.105.075128} {\bibfield
  {journal} {\bibinfo  {journal} {Phys. Rev. B}\ }\textbf {\bibinfo {volume}
  {105}},\ \bibinfo {pages} {075128} (\bibinfo {year} {2022})}\BibitemShut
  {NoStop}%
\bibitem [{\citenamefont {Bartlett}\ and\ \citenamefont
  {Zhao}(2023)}]{Bartlett2023}%
  \BibitemOpen
  \bibfield  {author} {\bibinfo {author} {\bibfnamefont {J.}~\bibnamefont
  {Bartlett}}\ and\ \bibinfo {author} {\bibfnamefont {E.}~\bibnamefont
  {Zhao}},\ }\href {https://doi.org/10.1103/PhysRevB.107.035101} {\bibfield
  {journal} {\bibinfo  {journal} {Phys. Rev. B}\ }\textbf {\bibinfo {volume}
  {107}},\ \bibinfo {pages} {035101} (\bibinfo {year} {2023})}\BibitemShut
  {NoStop}%
\bibitem [{\citenamefont {Jiang}\ and\ \citenamefont {Lee}()}]{Jiang2022}%
  \BibitemOpen
  \bibfield  {author} {\bibinfo {author} {\bibfnamefont {H.}~\bibnamefont
  {Jiang}}\ and\ \bibinfo {author} {\bibfnamefont {C.~H.}\ \bibnamefont
  {Lee}},\ }\href@noop {} {\bibinfo  {journal} {arXiv:2207.08843}\
  }\BibitemShut {NoStop}%
\bibitem [{\citenamefont {Kawabata}\ \emph {et~al.}(2019)\citenamefont
  {Kawabata}, \citenamefont {Shiozaki}, \citenamefont {Ueda},\ and\
  \citenamefont {Sato}}]{Kawabata2019}%
  \BibitemOpen
\bibfield  {journal} {  }\bibfield  {author} {\bibinfo {author} {\bibfnamefont
  {K.}~\bibnamefont {Kawabata}}, \bibinfo {author} {\bibfnamefont
  {K.}~\bibnamefont {Shiozaki}}, \bibinfo {author} {\bibfnamefont
  {M.}~\bibnamefont {Ueda}},\ and\ \bibinfo {author} {\bibfnamefont
  {M.}~\bibnamefont {Sato}},\ }\href
  {https://doi.org/10.1103/PhysRevX.9.041015} {\bibfield  {journal} {\bibinfo
  {journal} {Phys. Rev. X}\ }\textbf {\bibinfo {volume} {9}},\ \bibinfo {pages}
  {041015} (\bibinfo {year} {2019})}\BibitemShut {NoStop}%
\bibitem [{\citenamefont {Okugawa}\ \emph {et~al.}(2021)\citenamefont
  {Okugawa}, \citenamefont {Takahashi},\ and\ \citenamefont
  {Yokomizo}}]{Okugawa2021}%
  \BibitemOpen
  \bibfield  {author} {\bibinfo {author} {\bibfnamefont {R.}~\bibnamefont
  {Okugawa}}, \bibinfo {author} {\bibfnamefont {R.}~\bibnamefont {Takahashi}},\
  and\ \bibinfo {author} {\bibfnamefont {K.}~\bibnamefont {Yokomizo}},\ }\href
  {https://doi.org/10.1103/PhysRevB.103.205205} {\bibfield  {journal} {\bibinfo
   {journal} {Phys. Rev. B}\ }\textbf {\bibinfo {volume} {103}},\ \bibinfo
  {pages} {205205} (\bibinfo {year} {2021})}\BibitemShut {NoStop}%
\bibitem [{\citenamefont {Fukui}\ \emph {et~al.}(2005)\citenamefont {Fukui},
  \citenamefont {Hatsugai},\ and\ \citenamefont {Suzuki}}]{Fukui2005}%
  \BibitemOpen
  \bibfield  {author} {\bibinfo {author} {\bibfnamefont {T.}~\bibnamefont
  {Fukui}}, \bibinfo {author} {\bibfnamefont {Y.}~\bibnamefont {Hatsugai}},\
  and\ \bibinfo {author} {\bibfnamefont {H.}~\bibnamefont {Suzuki}},\
  }\href@noop {} {\bibfield  {journal} {\bibinfo  {journal} {J. Phys. Soc.
  Jpn.}\ }\textbf {\bibinfo {volume} {74}},\ \bibinfo {pages} {1674} (\bibinfo
  {year} {2005})}\BibitemShut {NoStop}%
\bibitem [{\citenamefont {Okugawa}\ \emph {et~al.}(2019)\citenamefont
  {Okugawa}, \citenamefont {Hayashi},\ and\ \citenamefont
  {Nakanishi}}]{Okugawa2019}%
  \BibitemOpen
  \bibfield  {author} {\bibinfo {author} {\bibfnamefont {R.}~\bibnamefont
  {Okugawa}}, \bibinfo {author} {\bibfnamefont {S.}~\bibnamefont {Hayashi}},\
  and\ \bibinfo {author} {\bibfnamefont {T.}~\bibnamefont {Nakanishi}},\ }\href
  {https://doi.org/10.1103/PhysRevB.100.235302} {\bibfield  {journal} {\bibinfo
   {journal} {Phys. Rev. B}\ }\textbf {\bibinfo {volume} {100}},\ \bibinfo
  {pages} {235302} (\bibinfo {year} {2019})}\BibitemShut {NoStop}%
\bibitem [{\citenamefont {Song}\ \emph
  {et~al.}(2020{\natexlab{b}})\citenamefont {Song}, \citenamefont {Liu},
  \citenamefont {Zheng}, \citenamefont {Zhang}, \citenamefont {Wang},\ and\
  \citenamefont {Lu}}]{YSong2020}%
  \BibitemOpen
  \bibfield  {author} {\bibinfo {author} {\bibfnamefont {Y.}~\bibnamefont
  {Song}}, \bibinfo {author} {\bibfnamefont {W.}~\bibnamefont {Liu}}, \bibinfo
  {author} {\bibfnamefont {L.}~\bibnamefont {Zheng}}, \bibinfo {author}
  {\bibfnamefont {Y.}~\bibnamefont {Zhang}}, \bibinfo {author} {\bibfnamefont
  {B.}~\bibnamefont {Wang}},\ and\ \bibinfo {author} {\bibfnamefont
  {P.}~\bibnamefont {Lu}},\ }\href
  {https://doi.org/10.1103/PhysRevApplied.14.064076} {\bibfield  {journal}
  {\bibinfo  {journal} {Phys. Rev. Applied}\ }\textbf {\bibinfo {volume}
  {14}},\ \bibinfo {pages} {064076} (\bibinfo {year}
  {2020}{\natexlab{b}})}\BibitemShut {NoStop}%
\bibitem [{\citenamefont {Yang}\ \emph {et~al.}(2020)\citenamefont {Yang},
  \citenamefont {Zhang}, \citenamefont {Fang},\ and\ \citenamefont
  {Hu}}]{Yang2020}%
  \BibitemOpen
  \bibfield  {author} {\bibinfo {author} {\bibfnamefont {Z.}~\bibnamefont
  {Yang}}, \bibinfo {author} {\bibfnamefont {K.}~\bibnamefont {Zhang}},
  \bibinfo {author} {\bibfnamefont {C.}~\bibnamefont {Fang}},\ and\ \bibinfo
  {author} {\bibfnamefont {J.}~\bibnamefont {Hu}},\ }\href
  {https://doi.org/10.1103/PhysRevLett.125.226402} {\bibfield  {journal}
  {\bibinfo  {journal} {Phys. Rev. Lett.}\ }\textbf {\bibinfo {volume} {125}},\
  \bibinfo {pages} {226402} (\bibinfo {year} {2020})}\BibitemShut {NoStop}%
\end{thebibliography}
\end{document}